\documentclass[useAMS,usenatbib]{mnras}

\usepackage{graphicx}
\usepackage{multirow}
\usepackage{tabularx}  
\usepackage{subcaption}
\usepackage{amssymb}
\usepackage{bold-extra}
\usepackage{ulem}

\newcommand{\rhocrit}{$\rho_\textrm{\small crit,0}$}

\def\pc{{\rm\thinspace pc}}
\def\kpc{{\rm\thinspace kpc}}

\def\Msun{\hbox{$\rm\thinspace M_{\odot}$}}

\def\yr{{\rm\thinspace yr}}

\def\Msunpc2{{\Msun\pc}^{-2}}
\def\Msunyrkpc2{{\Msun\yr^{-1}\kpc}^{-2}}

\def\magarcsec2{{\rm\thinspace mag\thinspace arcsec}^{-2}}

\title[Large-scale distribution of ionized metals]{The large-scale distribution of ionized metals in IllustrisTNG}
\author[M. C. Artale et al.]{M. Celeste Artale$^{1}$\thanks{Maria.Artale@uibk.ac.at, mcartale@gmail.com}, Markus Haider$^{1}$\thanks{mrkshdr@gmail.com},
Antonio D. Montero-Dorta$^{2,3}$\thanks{amonterodorta@gmail.com}, Mark Vogelsberger$^{4}$,
\newauthor
Davide Martizzi$^{5,6}$,  Paul Torrey$^{7}$, Simeon Bird$^{8}$, Lars Hernquist$^{9}$, \& Federico Marinacci$^{10}$ \\
$^{1}$Institut f\"ur Astro- und Teilchenphysik, Universit\"at Innsbruck, Technikerstra\ss e 25/8, A-6020 
Innsbruck, Austria\\
$^2$Departamento de F\'isica, Universidad T\'ecnica Federico Santa Mar\'ia, Casilla 110-V, Avda. Espa\~na 1680, Valpara\'iso, Chile\\
$^3$Departamento de F\'isica Matem\'atica, Instituto de F\'isica, Universidade de S\~ao Paulo, Rua do Mat\~ao 1371, CEP 05508-090, \\
S\~ao Paulo, Brazil \\
$^{4}$Kavli Institute for Astrophysics and Space Research, Massachusetts Institute of Technology, Cambridge, 
MA 02139, USA\\
$^{5}$Dark Cosmology Centre, Niels Bohr Institute, University of Copenhagen, 2100 Copenhagen, Denmark \\
$^{6}$Department of Astronomy and Astrophysics, University of California, Santa Cruz, CA 95064, USA \\
$^{7}$Department of Astronomy, University of Florida, 211 Bryant Space Sciences Center, Gainesville, FL 32611, USA \\
$^{8}$University of California Riverside, Department of Physics and Astronomy, Riverside, CA, 92521 \\
$^{9}$Harvard-Smithsonian Center for Astrophysics, 60 Garden St., Cambridge, MA 02138, USA\\
$^{10}$Department of Physics and Astronomy ``Augusto Righi", University of Bologna, via Gobetti 93/2, 40129 Bologna, Italy
}
\begin{document}

\date{}

\pagerange{\pageref{firstpage}--\pageref{lastpage}} \pubyear{2021}

\maketitle

\label{firstpage}

\begin{abstract} 
We study the intrinsic large-scale distribution and evolution of seven ionized metals in the IllustrisTNG magneto-hydrodynamical cosmological simulation. 
We focus on the fractions of C\,\textsc{ii}, C\,\textsc{iv}, Mg\,\textsc{ii}, N\,\textsc{v}, Ne\,\textsc{viii}, O\,\textsc{vi}, and Si\,\textsc{iv} in different cosmic web structures (filaments, haloes, and voids) and gas phases (warm-hot intergalactic medium  WHIM, hot, diffuse, and condensed gas) from $z=6$ to $z=0$. Our analysis provides a new perspective to the study of the distribution and evolution of baryons across cosmic time while offering new hints in the context of the well-known missing baryons problem. The cosmic web components are here identified using the local comoving dark matter density, which provides a simple but effective way of mapping baryons on large scales. 
Our results show that C\,\textsc{ii} and Mg\,\textsc{ii} are mostly located in condensed gas inside haloes in high-density and low-temperature star-forming regions 
($\rho_{\rm gas}/\bar{\rho}_{\rm bar}\gtrsim10^3$, and ${\rm T}\lesssim10^{5}$~K). 
C\,\textsc{iv} and Si\,\textsc{iv} present similar evolution of their mass fractions in haloes and filaments across cosmic time. In particular, their mass budgets in haloes in condensed phase ($\rho_{\rm gas}/\bar{\rho}_{\rm bar}\gtrsim10^3$, and ${\rm T}\lesssim10^{5}$~K) are driven by gas cooling and star formation with a peak at $z\sim2$.
Finally, our results confirm that O\,\textsc{vi}, Ne\,\textsc{viii}, and N\,\textsc{v} are good tracers of warm/hot and low-density gas  at low redshift ($\rho_{\rm gas}/\bar{\rho}_{\rm bar}\lesssim10^3$, and ${\rm T}\gtrsim10^{5}$~K), regions that are likely to contain most of the missing baryons in the local Universe.
\end{abstract}


\section{Introduction}

In the standard model of cosmology, the formation of the large scale structure of the Universe (LSS) is driven by the competing action of dark matter and dark energy. During this process, baryons undergo gravitational collapse and condense to form the visible web of filaments and knots, separated by void regions, that we call the {\it{cosmic web}}. At high redshift, the amount of baryons measured from the Cosmic Microwave Background and Ly$\alpha$ forest, or derived from Big Bang nucleosynthesis is in good agreement with theoretical expectations. At low redshift, however, observational studies report systematically lower values, with around 30 percent of the baryons unaccounted for. The so called \textit{missing baryon problem} refers to this apparent lack of baryons as measured from observations \citep{Kirkman2003,Bregman2007,Nicastro2008,Pettini1999,Anderson2010,Shull2012,Nicastro2017}, which might be due to the fact that a large amount of gas lies in low-density, high-temperature regions that are hard to detect using current UV and X-ray instrumentation.

Baryons are, in fact, also missing in galaxies, where the baryonic fraction is much lower than the universal baryon-to-total-mass ratio \citep{Bregman2007,McGaugh2010}. The difference is more pronounced for smaller galaxies, suggesting that haloes with shallower potential wells are less efficient in terms of retaining baryons. Overall, these results seem to indicate that the ``missing baryons" could be somewhere in the circum-galactic space and possibly beyond the virial radius of the haloes
as well, as suggested by recent reports \citep[see e.g.,][]{Tumlinson2013,Bordoloi2014,Werk2014,Johnson2015,Johnson2017,Tumlinson2017}.

The baryons missing in galaxies and in the LSS are indeed part of the same problem. Theoretical models predict that a large fraction of them resides in filamentary structure conforming to the so-called warm-hot intergalactic medium (WHIM), composed of hot (T~$=10^5 - 10^7$~K) low-density gas ($n_b = 10^{-6} - 10^{-4} {\rm cm}^{-3}$). These baryons are initially shock-heated during the build-up of the large-scale structures. Subsequently, stellar feedback and galactic winds act as additional heating sources, regulating also the chemical abundances in the WHIM \citep{Cen1999,Cen2006,Dave2001}.   

The gas in the WHIM can be detected through the radiation field it produces (i.e., emission lines), or through absorption lines in background sources. Due to its low density, current X-ray telescopes fail to detect most of the WHIM emission. Nevertheless, several reports show that
filamentary and intra-cluster structures can be mapped from the 
tail of the WHIM X-ray intensity distribution \citep{Scharf2000,Markevitch2003,Kaastra2003,Eckert2015,Hattori2017,Connor2018,Tanimura2020}. These types of measurements are expected to be possible in forthcoming years, thanks to, e.g., the Athena space mission \citep[][]{ATHENA}.

The most common way to trace the WHIM is through absorption lines in UV/X-ray quasar spectra. Significant evidence of unaccounted baryons has been reported from O\,\textsc{vi} absorption lines measured with the Far Ultraviolet Spectroscopic Explorer (FUSE) and the Hubble Space Telescope (HST-COS) \citep{Savage1998,Tripp2000,Danforth2016}. 
Measurements based on O\,\textsc{vii} and O\,\textsc{viii} absorption lines from Chandra and XMM-Newton have also been presented \citep{Fang2002,McKernan2003}. Several other lines have been used in similar analyses, including C\,\textsc{ii}, C\,\textsc{iv}, N\,\textsc{v}, Si\,\textsc{iii}, Si\,\textsc{iv}, and Ne\,\textsc{viii} \citep[see e.g.,][]{Cooksey2010,Nicastro2013,Danforth2016,Tejos2016,Burchett2019,Manuwal2019,Ahoranta2020,Chen2020}. The main disadvantage of absorption lines is that they can only provide information about the gas along the line of sight of the background quasar. Despite this limitation, the absorption-line approach has proven successful, as recently confirmed by \citet{Nicastro2018}  \citep[see also,][]{Nicastro2018_b}. Based on measurements of O\,\textsc{vii} absorption lines in the X-ray spectra of a blazar at $z=0.48$, the authors claim the detection of a significant fraction of the missing baryons in the WHIM. 

An alternative way to investigate the gas in the WHIM is through the Sunyaev-Zel'dovich effect \citep[SZE,][]{Sunyaev1972,Mroczkowski2019}. Recent works have shown the potential of the SZE in terms of constraining the distribution and abundance of gas in the WHIM \citep{Hernandez2015,Hill2016,deGraaff2019,Tanimura2019,Lim2020,Chaves-Montero2021}. In this context, future instruments such as CMB-S4 \citep{Abazajian2016} are expected to be crucial to shed light onto the missing baryon problem.

From a theoretical perspective, hydrodynamical cosmological simulations have proven to be a useful and reliable tool to investigate the location and distribution of baryons in the cosmic web \citep{Cen1999,Dave2001,Cen2006,Shull2012,Roca2016,Suresh2017,GarragaEspinosa2021}. In fact, several analyses based on hydrodynamical simulations predict that the missing baryons are in the form of warm-hot diffuse gas located between galaxies in the circumgalactic medium (CGM) and beyond the virial radius in the intergalactic medium (IGM). Moreover, these simulations have shown that star formation regulates the production rate of metals, while processes such as supernova feedback, galactic winds, and active galactic nucleus (AGN) feedback are capable of expelling the enriched hot gas into the CGM and IGM \citep{Cen2006,Theuns2002,Oppenheimer2006,Cen2011,Tescari2011,Rahmati2016,Wijers2020}. In this intricate scenario, the impact of metals on the cooling properties of gas must also be considered.

In this work, we use the IllustrisTNG hydrodynamical cosmological simulation suite \citep{Pillepich2018b,Pillepich2018,Nelson2018_ColorBim,Nelson2018_Oxigen,Nelson2019,Marinacci2018,Naiman2018,Springel2018}
to investigate the evolution and distribution of ionized metals across cosmic time, between $z=0-6$. Our main goal is to provide a theoretical analysis to track the evolution of the different ionized metals in time. IllustrisTNG has proven capable of reproducing different observational features of galaxy formation and evolution, such as the evolution of the galaxy mass-metallicity relation, the galaxy color bimodality, and the cosmic star formation rate, among others. In particular, \citet{Nelson2018_Oxigen} use this simulation to investigate the spatial distribution and physical properties of O\,\textsc{vi}, O\,\textsc{vii}, and O\,\textsc{viii} in the CGM and IGM. Their results show that IllustrisTNG achieves a good level of agreement with observations of the column density distribution function for O\,\textsc{vi} at low redshift \citep{Danforth2008,Thom2008,Tripp2008,Danforth2016}. 
\citet{Nelson2018_Oxigen} also explore the impact of AGN and stellar feedback on the physical state of the CGM. Their results suggest that these mechanisms are fundamental drivers of the properties of the CGM. In particular, the low-accretion AGN feedback is crucial to explaining the differences in the N$_{{\rm O}\,\textsc{vi}}$ columns between  star-forming and quiescent galaxies.

Our analysis focuses on a selected set of seven ions (C\,\textsc{ii}, C\,\textsc{iv}, Mg\,\textsc{ii}, N\,\textsc{v}, Ne\,\textsc{viii}, O\,\textsc{vi}, and Si\,\textsc{iv}), which are chosen because they are the ones more commonly detected in observations\footnote{In particular, absorption lines from C\,\textsc{ii}, C\,\textsc{iv}, O\,\textsc{vi}, Mg\,\textsc{ii}, and Si\,\textsc{iv} are commonly detected by quasar spectra.}. For this set of ions, we study their distribution in the cosmic web (filaments, knots and voids) along with their abundances in the different gas phases across cosmic time. Our procedure follows that of \citet{Haider2016} but it is applied to a more sophisticated hydrodynamical simulation.  

The structure of the paper is the following. In \S~\ref{sec:SimuAndMetods} we provide a summary of the main properties of the IllustrisTNG simulations (\ref{sec:simu}), and the methodology adopted to identify the cosmic web (\ref{sec:definition_cosmicWeb}), and gas phases (\ref{sec:definition_ions}).
The results are presented in \S~\ref{sec:results}. Finally, the summary and conclusions are presented in \S~\ref{sec:conclusions}.  

\section{Simulation \& Methods}\label{sec:SimuAndMetods}
\subsection{Simulation data}\label{sec:simu}

The main results of this paper are based on the IllustrisTNG\footnote{http://www.tng-project.org} project, whereas the previous Illustris simulation is used for comparison. IllustrisTNG is a suite of magneto-hydrodynamical cosmological simulations that model the formation and evolution of galaxies within the standard $\Lambda$-CDM paradigm \citep{Pillepich2018b,Pillepich2018,Nelson2018_ColorBim,Nelson2018_Oxigen,Nelson2019,Marinacci2018,Naiman2018,Springel2018}.
IllustrisTNG is performed with the moving-mesh {\sc arepo} code \citep{Springel2010}
and based on the sub-grid models implemented on the previous Illustris simulation \citep{Vogelsberger2014,Genel2014,Sijacki2015}.

The main improvements provided by IllustrisTNG as compared to its predecessor are the inclusion of magneto-hydrodynamics, an updated scheme for galactic winds, and a new kinetic black hole feedback model for the low accretion state \citep[see,][for further details]{Pillepich2018,Weinberger2017}. 
IllustrisTNG includes sub-grid models that accounts for physical processes in galaxies such as star-formation, 
metal-line gas cooling, stellar feedback from supernovae Type Ia, II and asymptotic giant branch stars (AGB),
and AGN feedback. 
The model also follows the production and evolution of the following nine elements: H, He, C, N, O, Ne, Mg, Si, and Fe. 
The model includes an update of the tabulated stellar yields
for the different stellar feedback channels \citep[see][for further  details]{Naiman2018,Pillepich2018}. 
The sub-grid parameters of IllustrisTNG were calibrated to reproduce the cosmic star formation density, the galaxy stellar mass function at present, and the stellar-to-halo mass relation at $z=0$ \citep{Pillepich2018}.

The IllustrisTNG suite consists of simulations with three different box sizes publicly available on the website \citep{Nelson2019}. In this work, we use
two of the boxes: the ones with 75~$h^{-1}$~Mpc~(=~110.7~Mpc) and 205~$h^{-1}$~Mpc~(=~302.6~Mpc) side lengths (hereafter TNG100 and TNG300, respectively). Both boxes were run from $z=127$ up to $z=0$.
In TNG100, dark matter particles and gas cells are modeled initially through 1820$^3$ elements each, with masses of
$7.46\times10^{6}~{\rm M}_{\odot}$ and $1.39\times10^{6}~{\rm M}_{\odot}$, respectively.
The TNG300 run, conversely, follows dark matter and gas using 2500$^3$ elements each, with masses of 
$5.88\times10^{7}~{\rm M}_{\odot}$ and $1.1\times10^{7}~{\rm M}_{\odot}$, respectively.  
The gravitational softening lengths for dark matter and stars is of 1.0~$h^{-1}$~kpc for TNG300
and 0.5~$h^{-1}$~kpc for TNG100. The
IllustrisTNG suite was run using the  \cite{Planck2016} cosmology, defined by the following cosmological parameters:
$\Omega_{\rm m} = 0.3089$, $\Omega_{\rm b} = 0.0486$, $\Omega_{\Lambda}=0.6911$, ${\rm H}_{0} = 100~h~{\rm km~s^{-1}~Mpc^{-1}}$,
$h=0.6774$, spectral index $n_{s} = 0.9667$ and normalization $\sigma_{8} = 0.8159$.

 The IllustrisTNG suite presents excellent agreement with important observational constraints such as the galaxy color bimodality at low redshift \citep{Nelson2018_ColorBim},
the galaxy clustering at $z=0.1$ \citep{Springel2018}, the stellar mass content of galaxy clusters \citep{Pillepich2018b}, the black hole -- stellar mass relation \citep{Weinberger2018}, and the mass-metallicity evolution of galaxies \citep{Torrey2019}.

In the more specific context of this work, IllustrisTNG 
has shown good agreement with several CGM and IGM properties reported from observations, which underpins the scientific relevance of our analysis. \citet{Nelson2018_Oxigen}, prove that the OVI content in the CGM and IGM is in good agreement with COS-Halos \citep[][]{Tumlinson2011}. Also, \citet{Truong2020} perform a complete analysis of the X-ray emission from diffuse, hot, metal-enriched gas with TNG50 and TNG100. Their results demonstrate that the IllustrisTNG suite is consistent with observed X-ray luminosity emitted by late- and early-type galaxies.
Other works such as \citet{Davies2020} and \citet{Terrazas2020} explore gas properties in the CGM, suggesting a strong correlation with the specific star formation rate of the galaxies. 
\citet{Nelson2019} investigates the outflows and dynamics of gas, whereas \citet{Nelson2020,Nelson2021} focus on the cold gas in the CGM through MgII covering fractions in luminous red galaxies.
Finally, \citet{Torrey2019} prove that the IllustrisTNG suite includes a comprehensive feedback model that widely distributes the metal budget into different gas phases.
All these results make us confident that the IllustrisTNG simulation is suitable for the goal of the present work.

In the first part, we make use of the gravitationally bound dark matter haloes from each simulation. In the Illustris and IllustrisTNG simulations, haloes are identified using a friends-of-friends algorithm (FOF) that adopts a linking length of 0.2 times the mean inter-particle separation \citep{Davis1985}. 
We characterize the mass of the dark matter haloes using M$_{\rm 500c}$, defined as the mass enclosed within a sphere of radius R$_{\rm 500c}$ (i.e., the radius at which the enclosed density equals 500 times the critical density). The total baryonic mass of each halo is also computed within a sphere of radius R$_{\rm 500c}$, referred to as M$_{\rm 500,baryon}$.

\subsection{Cosmic-web and gas-phase classifications}\label{sec:definition_cosmicWeb}

%%%%%%%%%%%%%%%%%%%%%%%%%%%
%%%%%%%%%%%%%%%%%%%%%%%%%%%
%%%%%%%%%  FIG 1  %%%%%%%%%%
%/n/home01/mcartale/MARKUS_FOLDERS/illustrisTNG/partitions
\begin{figure*}
\centering

 \large{$z=0$}
  \medskip
  
 \begin{subfigure}{0.24\linewidth}
 \includegraphics[width=\textwidth]{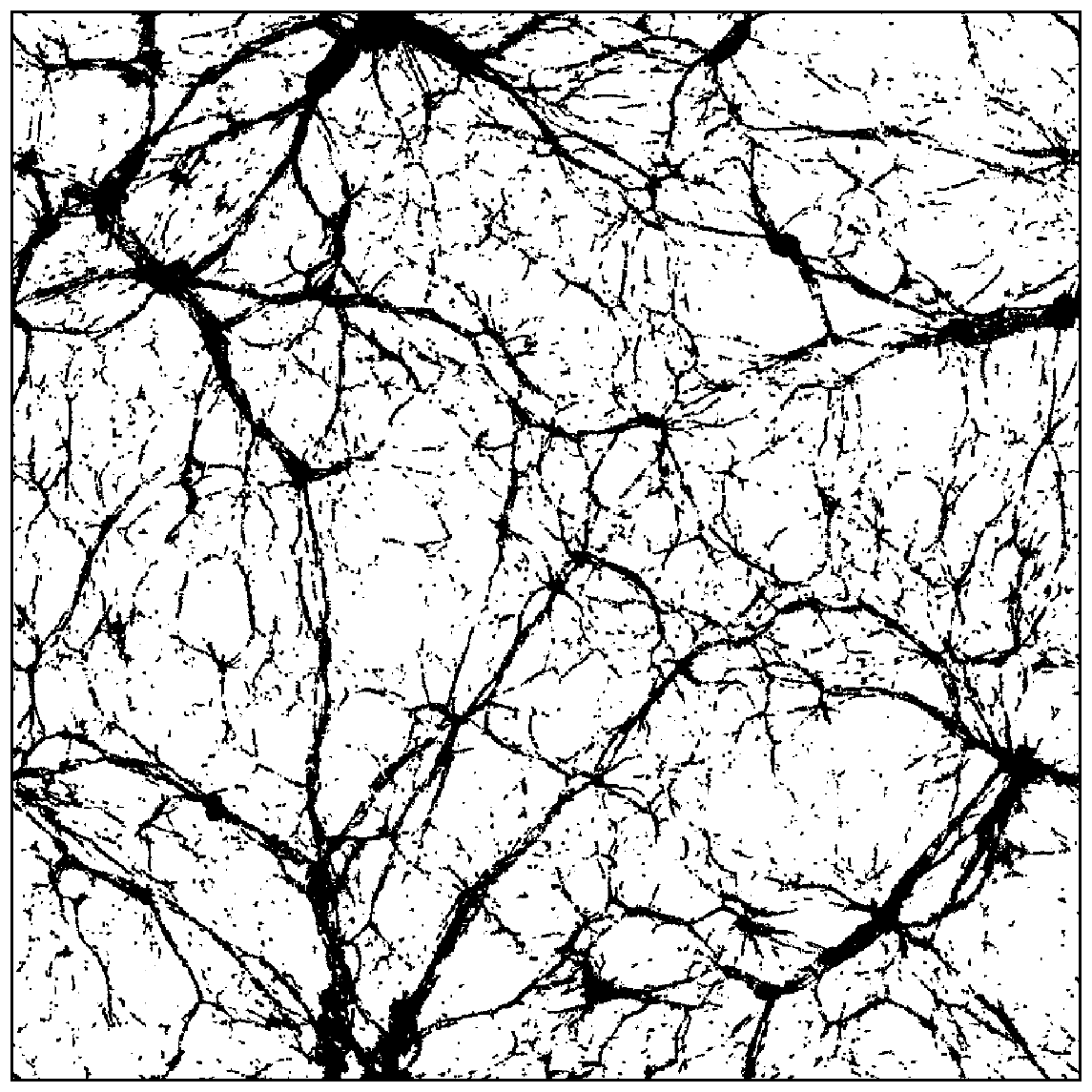}
 \caption*{Voids ($< 0.1 \rho_{\rm crit,0}$)}\label{fig:distr_voids}
 \end{subfigure}
 \begin{subfigure}{0.24\linewidth}
 \includegraphics[width=\textwidth]{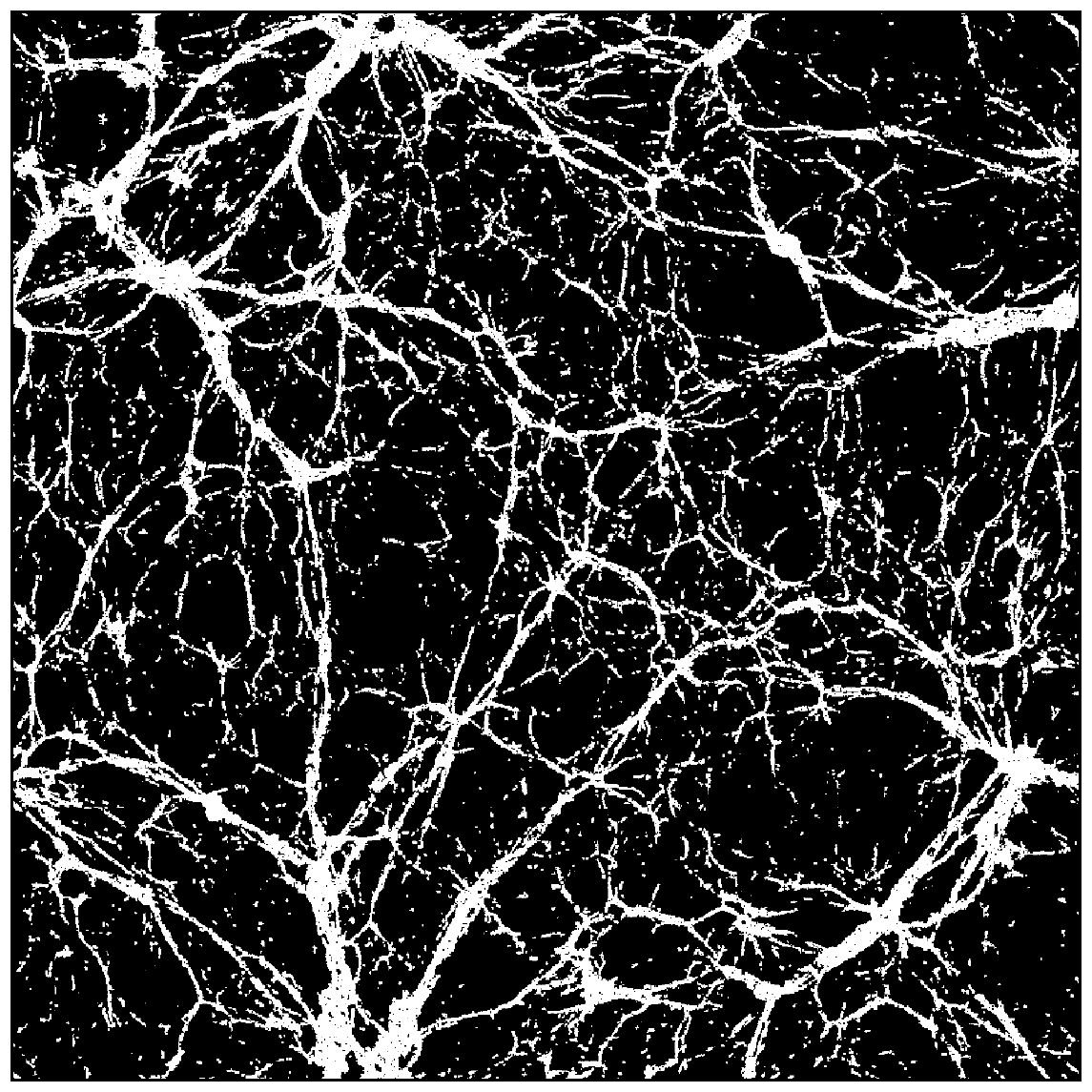}
 \caption*{Filaments ($0.1-57 \rho_{\rm crit,0}$)}\label{fig:distr_filaments}
 \end{subfigure}
 \begin{subfigure}{0.24\linewidth}
 \includegraphics[width=\textwidth]{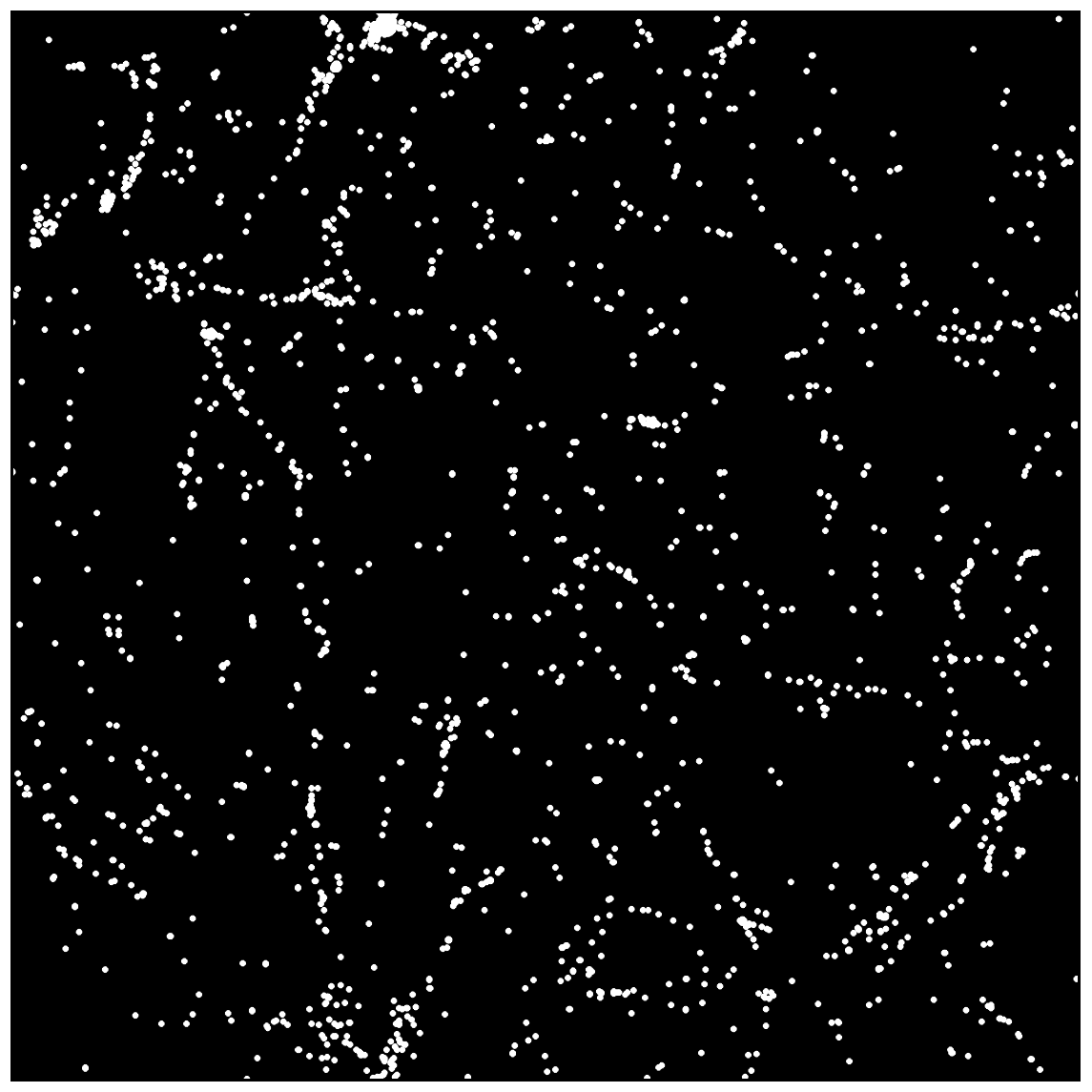}
 \caption*{Haloes ($> 57 \rho_{\rm crit,0}$)}\label{fig:distr_haloes}
 \end{subfigure}
 \begin{subfigure}{0.24\linewidth}
 \includegraphics[width=\textwidth]{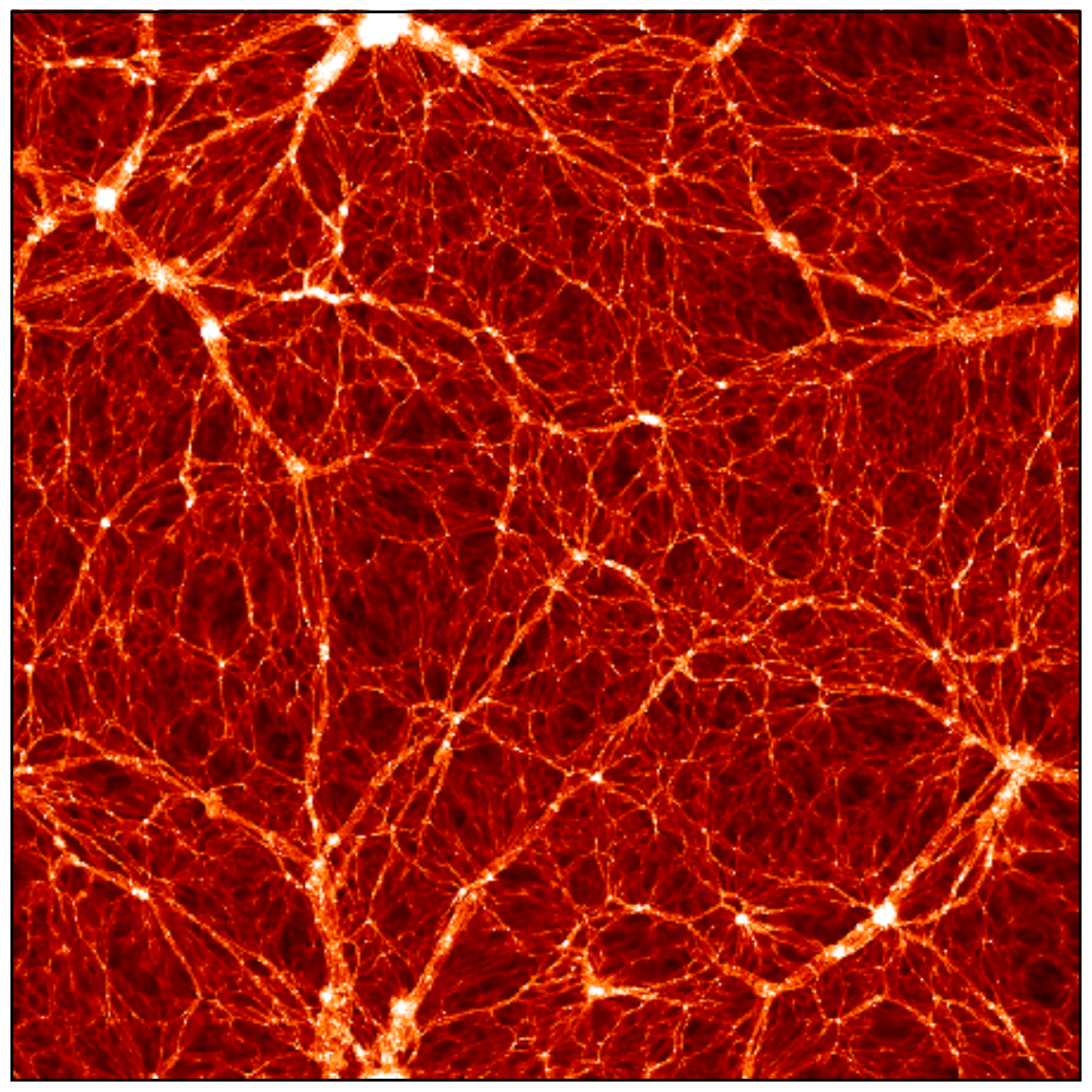}
 \caption*{Dark matter density}\label{fig:distr_dens}
 \end{subfigure}
 
  \medskip
  \medskip
 
 \large{$z=2$}
  \medskip

 \begin{subfigure}{0.24\linewidth}
 \includegraphics[width=\textwidth]{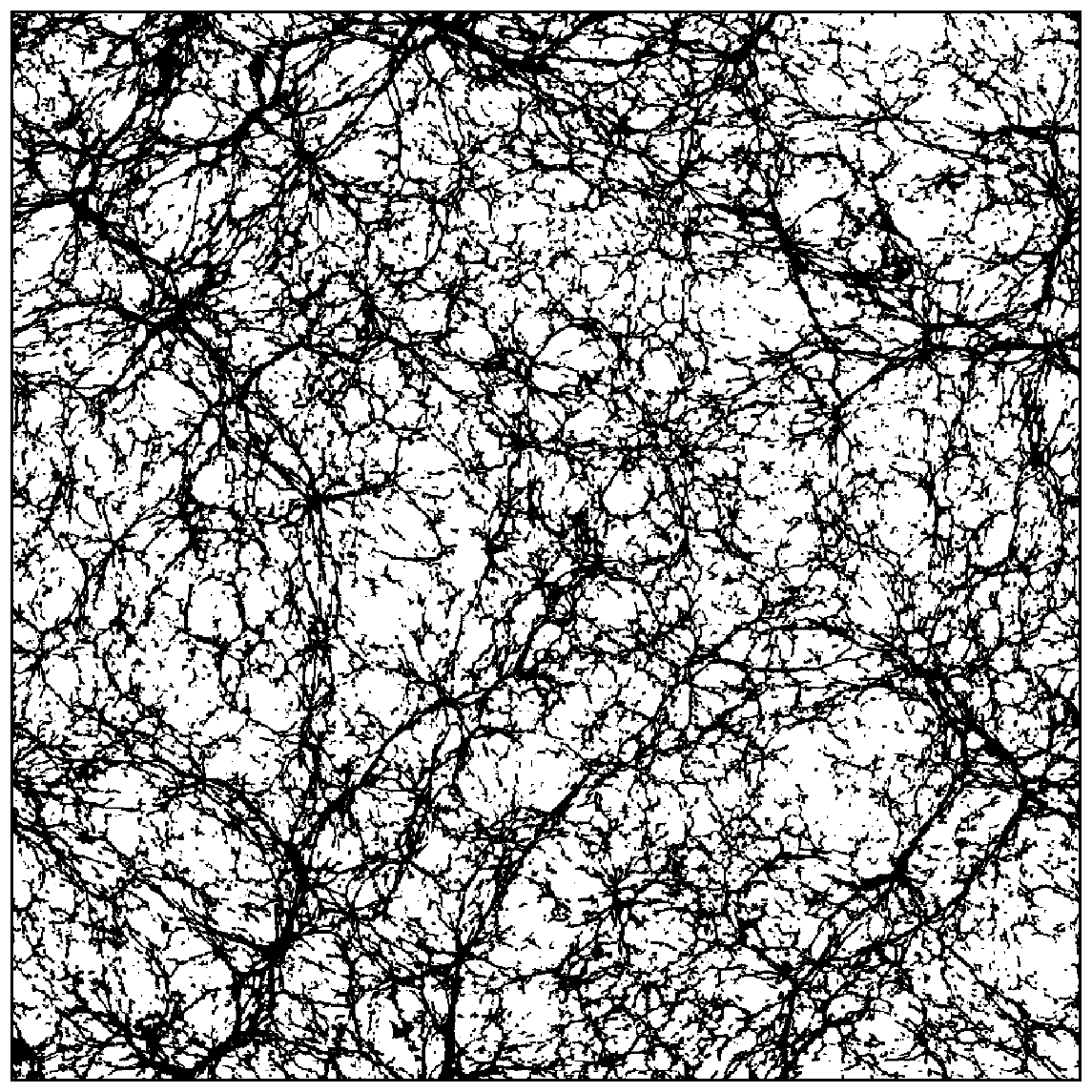}
 \caption*{Voids ($< 0.1 \rho_{\rm crit,0}$)}%\label{fig:distr_voids}
 \end{subfigure}
 \begin{subfigure}{0.24\linewidth}
 \includegraphics[width=\textwidth]{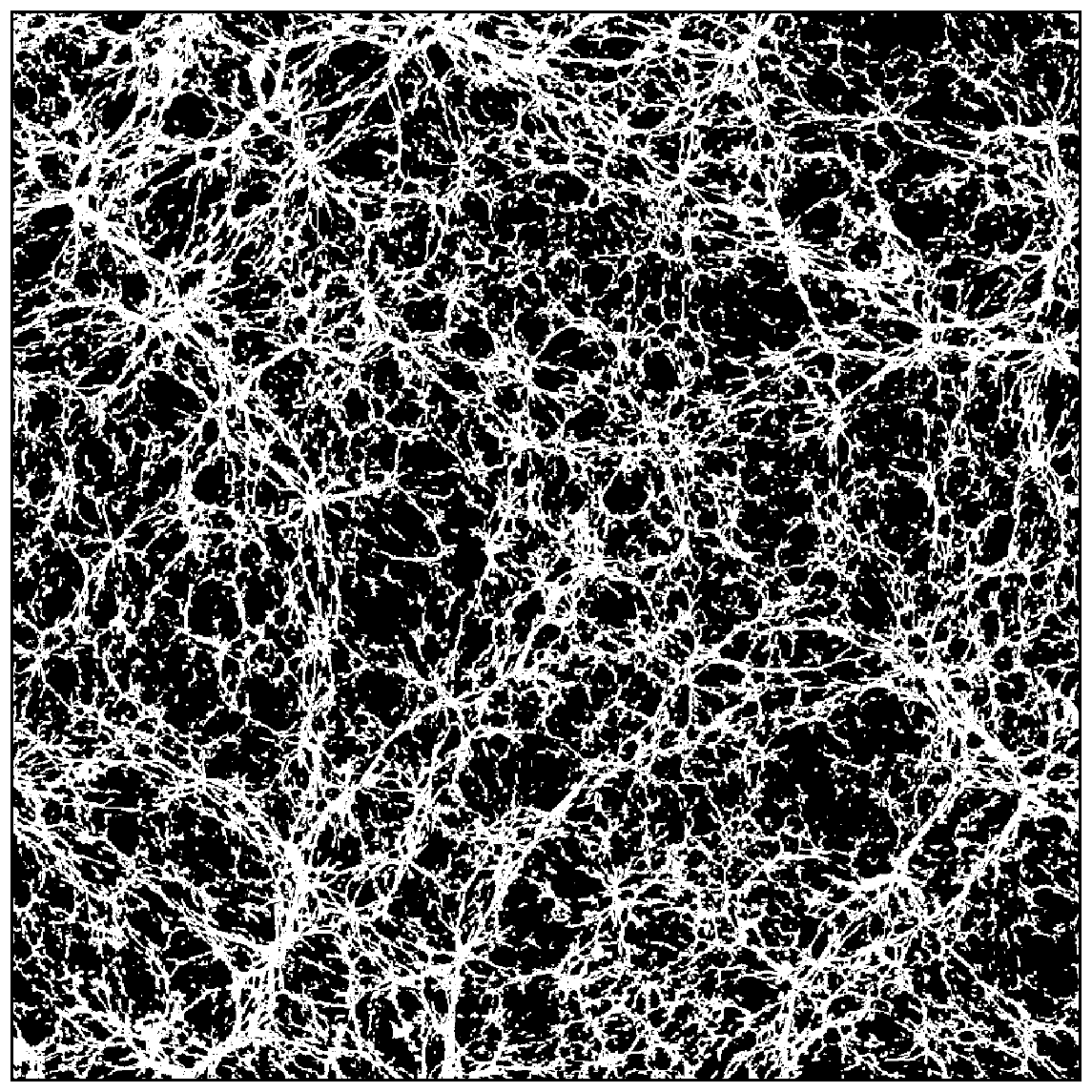}
 \caption*{Filaments ($0.1-57 \rho_{\rm crit,0}$)}%\label{fig:distr_filaments}
 \end{subfigure}
 \begin{subfigure}{0.24\linewidth}
 \includegraphics[width=\textwidth]{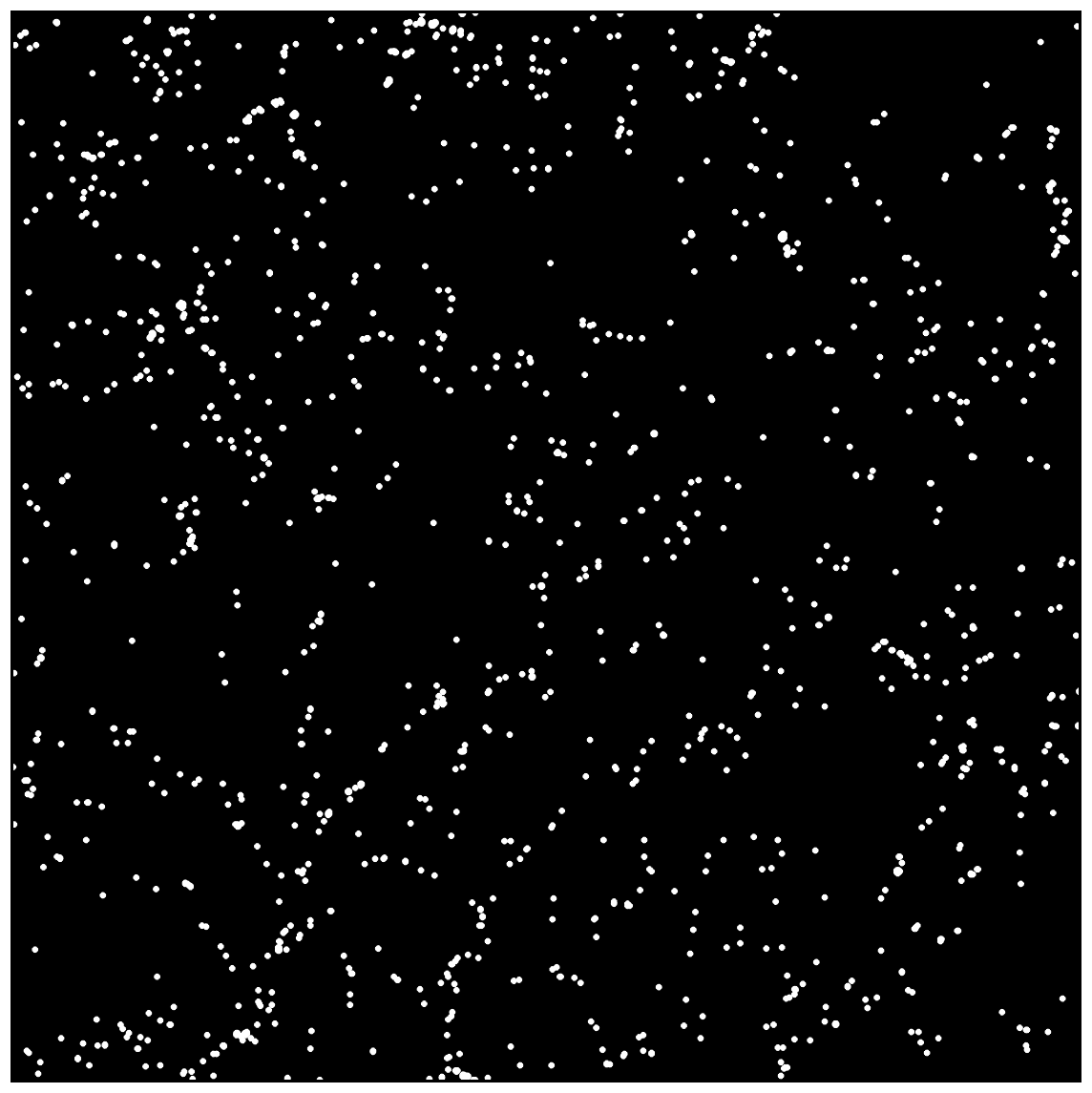}
 \caption*{Haloes ($> 57 \rho_{\rm crit,0}$)}%\label{fig:distr_haloes}
 \end{subfigure}
 \begin{subfigure}{0.24\linewidth}
 \includegraphics[width=\textwidth]{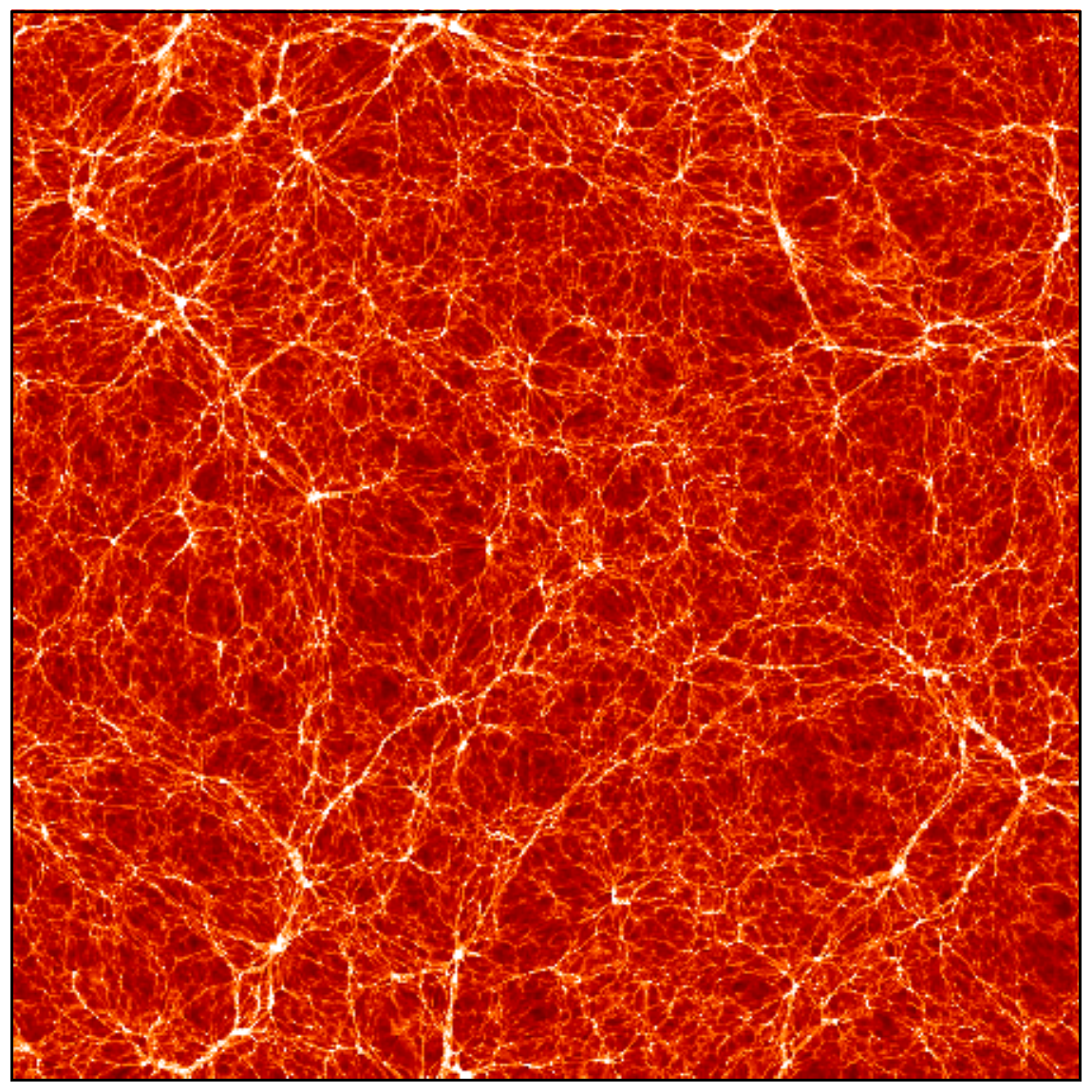}
 \caption*{Dark matter density}%\label{fig:distr_dens}
 \end{subfigure}
 
  \medskip
  \medskip

 \large{$z=6$}
   \medskip

 \begin{subfigure}{0.24\linewidth}
 \includegraphics[width=\textwidth]{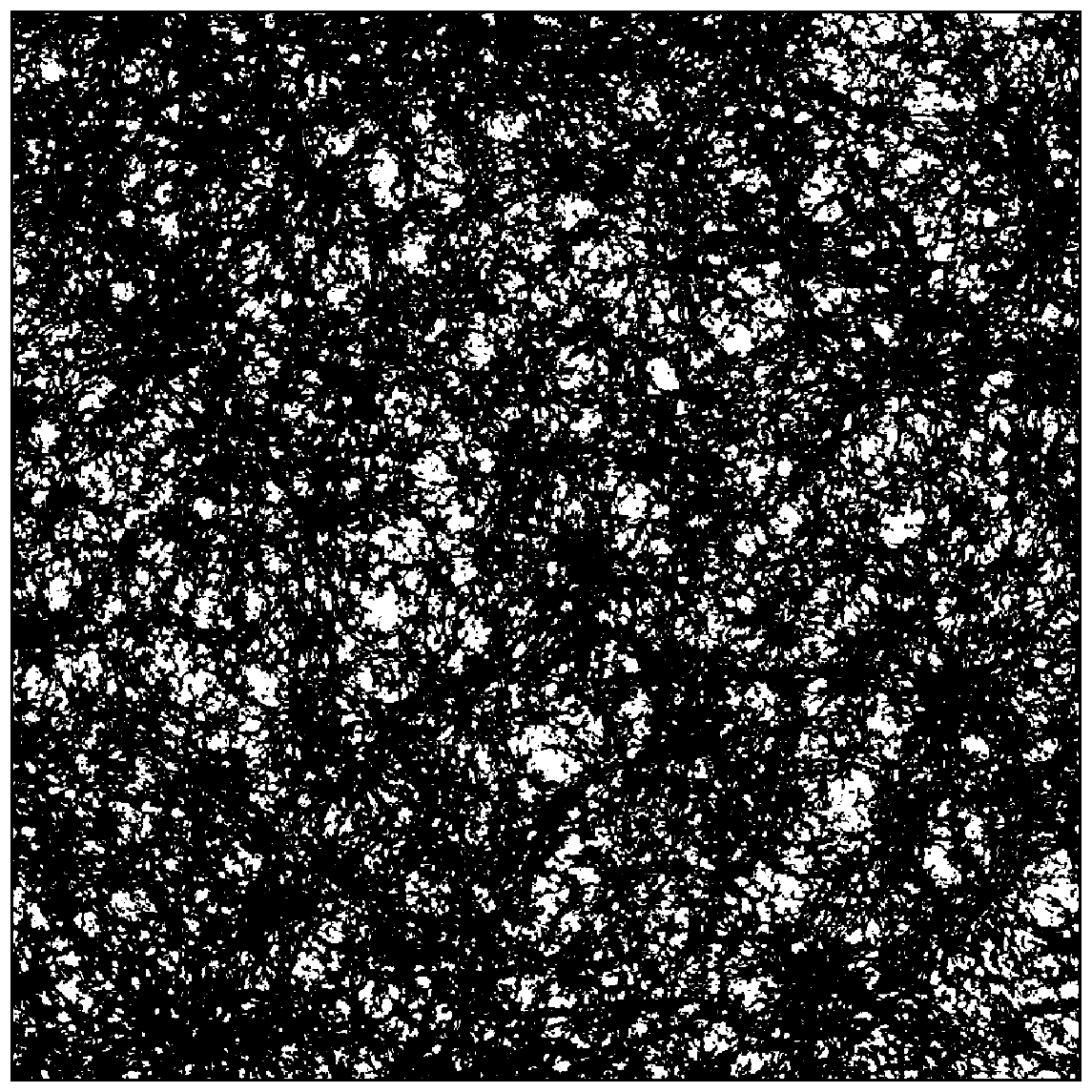}
 \caption*{Voids ($< 0.1 \rho_{\rm crit,0}$)}\label{fig:distr_voids}
 \end{subfigure}
 \begin{subfigure}{0.24\linewidth}
 \includegraphics[width=\textwidth]{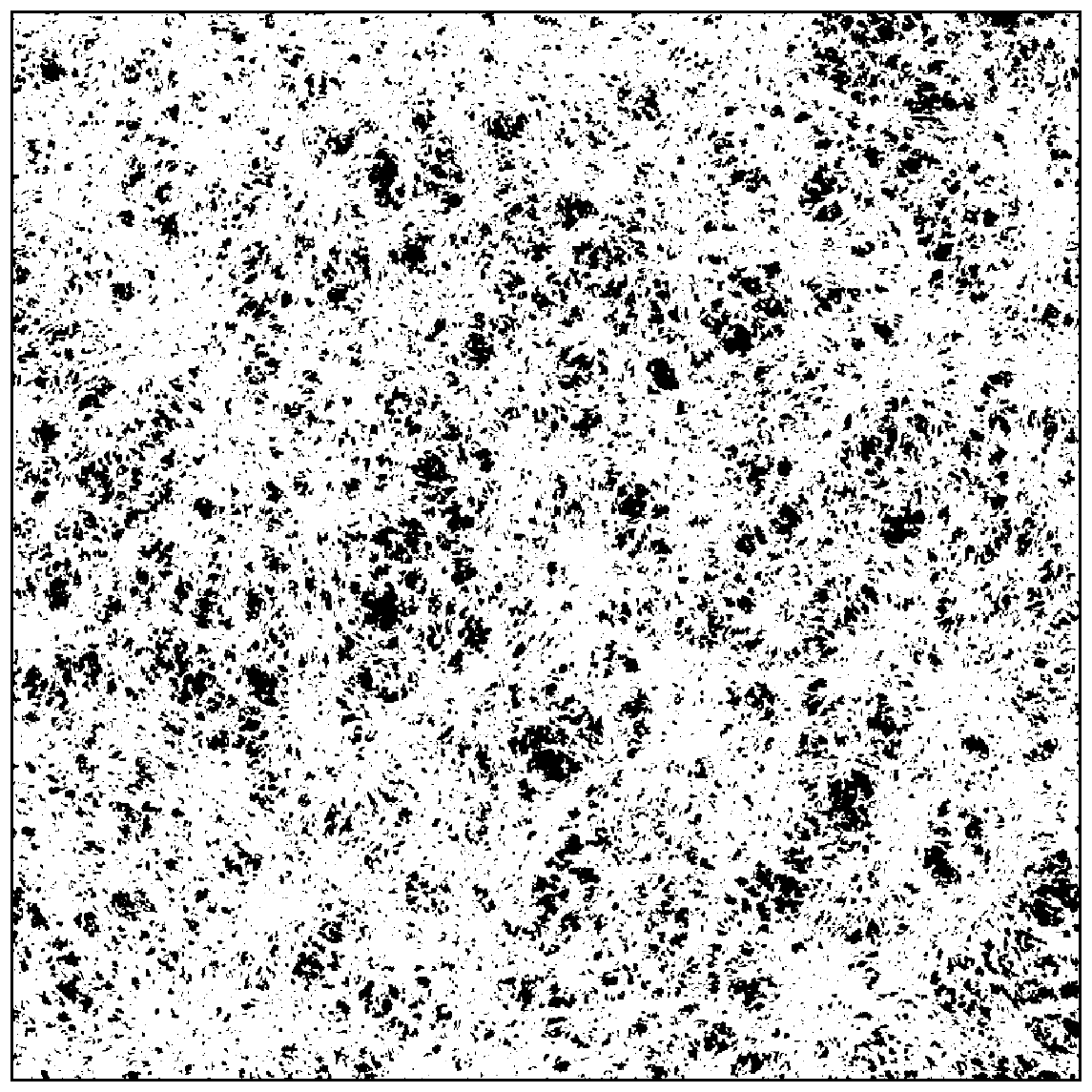}
 \caption*{Filaments ($0.1-57 \rho_{\rm crit,0}$)}\label{fig:distr_filaments}
 \end{subfigure}
 \begin{subfigure}{0.24\linewidth}
 \includegraphics[width=\textwidth]{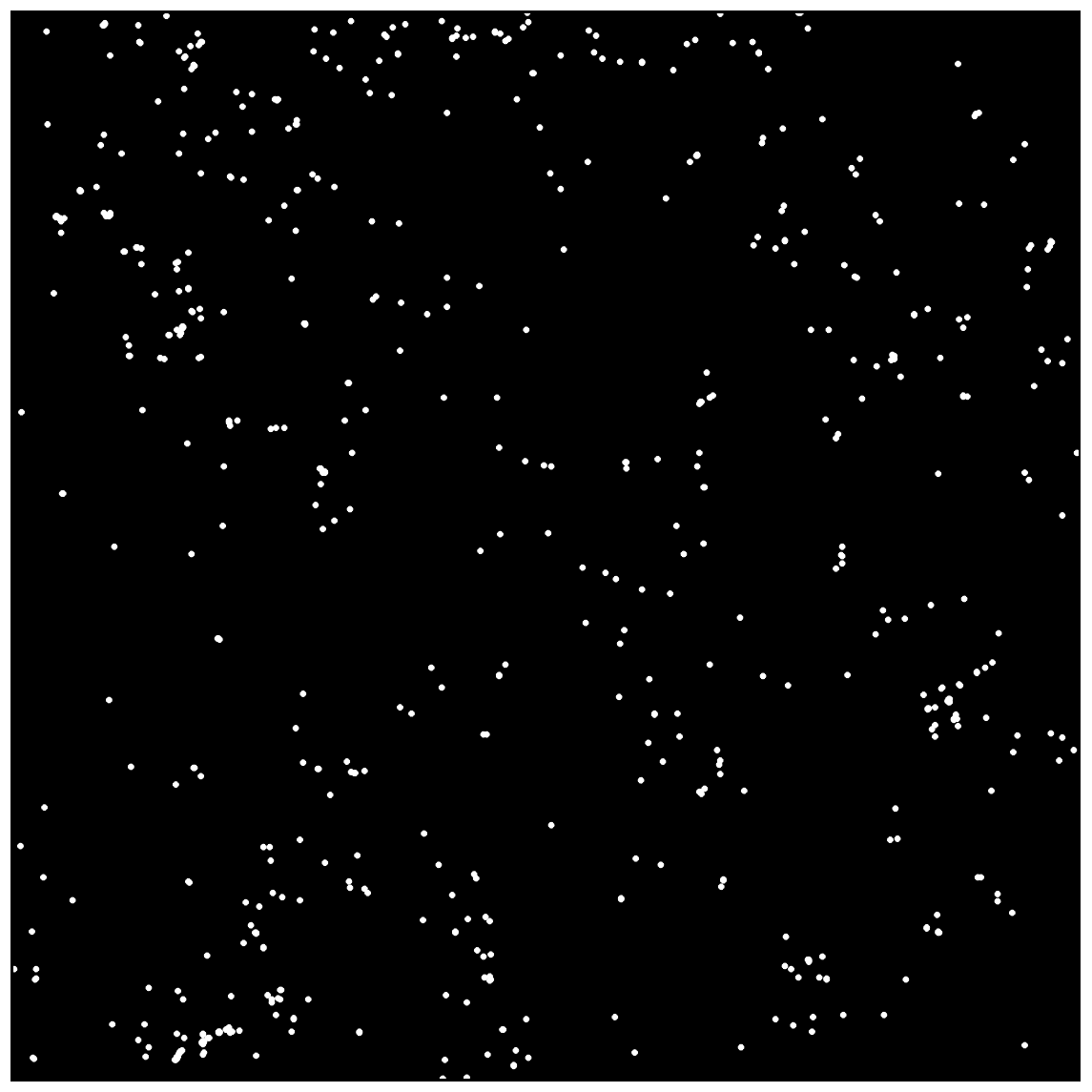}
 \caption*{Haloes ($> 57 \rho_{\rm crit,0}$)}\label{fig:distr_haloes}
 \end{subfigure}
 \begin{subfigure}{0.24\linewidth}
 \includegraphics[width=\textwidth]{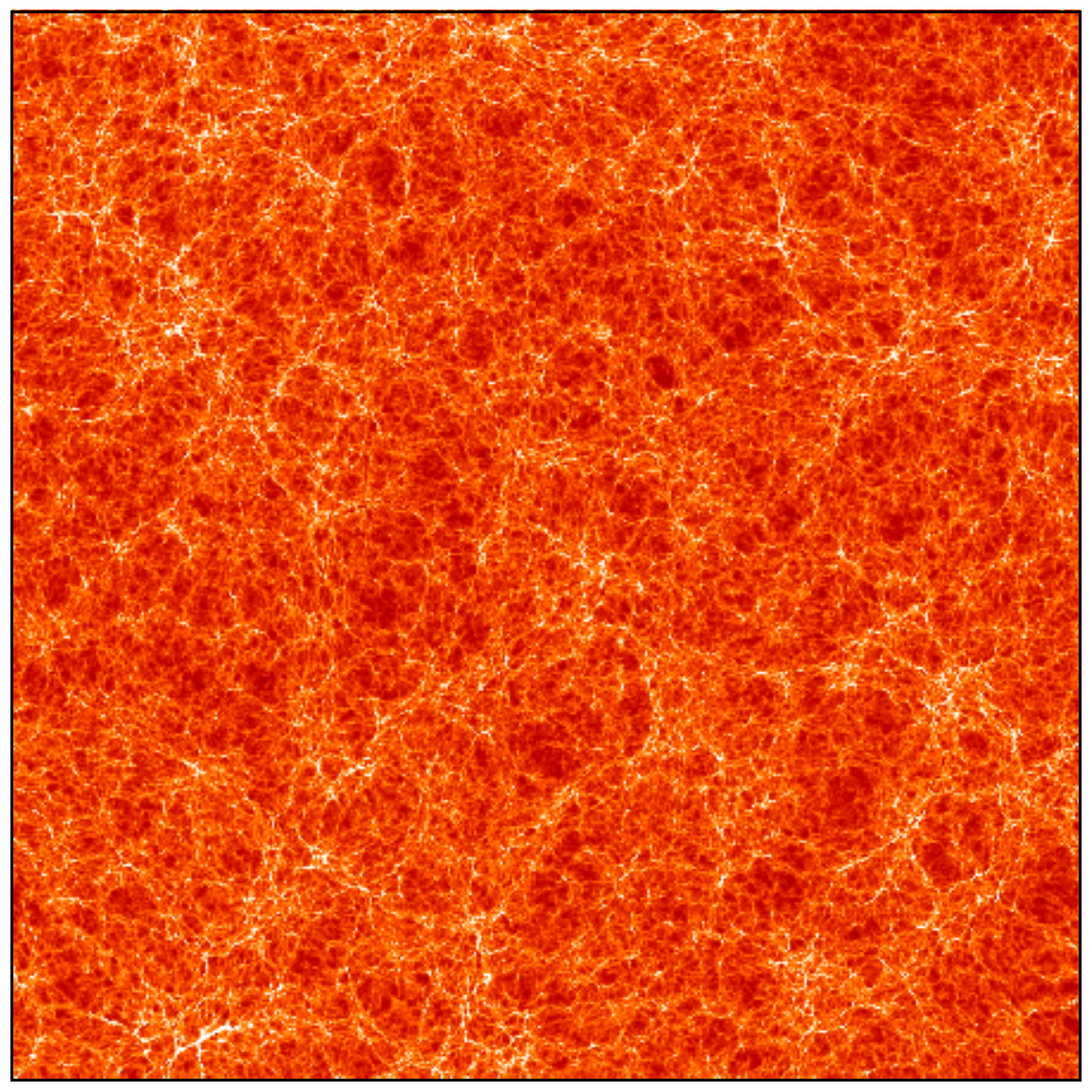}
 \caption*{Dark matter density}\label{fig:distr_dens}
 \end{subfigure}
\caption{Spatial distribution of the different cosmic-web regions (voids, filaments, and haloes/knots) for a slice of TNG100 at $z=0$, 2, and 6  (top, middle, and bottom panel, respectively). The cells corresponding to each region are drawn in white, otherwise they are in black. The haloes are highlighted with larger white points to make the figure more clear.  
The rightmost panel shows the local comoving dark matter density to facilitate the comparison. }\label{fig:regions}
\end{figure*}
%%%%%%%%%%%%%%%%%%%%%%%%%%%
%%%%%%%%%%%%%%%%%%%%%%%%%%%
%%%%%%%%%%%%%%%%%%%%%%%%%%%

We use the comoving local dark matter density to identify different cosmic-web environments in the simulations, following the methodology of \citet{Haider2016}\footnote{Our method differs only slightly from that of \citet{Haider2016}. While they use the average dark matter densities in cells over a volume of (104 kpc)$^{3}$ each, we implement the local dark matter density available in the IllustrisTNG database, computed in spheres with $\sim 3-60 {\rm kpc}$ of diameter at $z=0$.}. In this work, the comoving local dark matter density ($\rho_{dm}$) of each gas cell is estimated as the standard cubic-spline SPH kernel over a radius enclosing 64$\pm$4 dark matter particles\footnote{SubfindDMDensity, on the IllustrisTNG database.}. Our criteria to define environments is the following:
\begin{itemize}
    \item \textit{Haloes} (or knots) are regions where $\rho_{dm}~>~57~\rho_{\rm crit,0}$. This threshold corresponds approximately with the local density of a dark matter halo at its virial radius\footnote{The virial radius, R$_{200}$, is defined as the radius enclosing a region with density equal to 200 times the critical density.}, assuming a Navarro-Frenk-White density profile \citep[][]{Navarro1997}. 
    \item \textit{Filaments} are defined as regions with local dark matter density $\rho_{dm} = 0.1 - 57$~$\rho_{\rm crit,0}$. This definition encloses also the concept of \textit{sheets} (see Appendix~\ref{fig:Appendix_CosmicStructures} for further details).
    \item \textit{Voids} correspond to regions with $\rho_{dm}<0.1$~$\rho_{\rm crit,0}$.
\end{itemize}

 Figure~\ref{fig:regions} illustrates the criteria adopted to identify the cosmic web components in a slice of the TNG100 box at $z=0$, 2, and 6. Voids, filaments, and haloes are shown in white. The right panel shows the local dark matter density for the same slice.  
 The efficiency of our method in terms of tracing cosmic web structures have been shown in several other previous analyses \citep{Sousbie2011,Cautun2013,Cautun2014}. Note that this approach yields a cosmic web classification in agreement with that derived from tidal fields, for redshifts $z \lesssim 4$, as shown in \citet{Martizzi2018}. As stated by the authors, the main difference between these two approaches is the smoothing scales adopted. While our method employs a density field smoothed by the local SPH kernel (i.e., spheres of radius $\sim 1 - 20$~kpc/$h$ at $z=0$), the method based on the deformation tensor uses larger smoothing scales, of typically 2--6~Mpc/$h$. As a consequence, the local dark matter density field method is more subject to small-scale fluctuations.
As shown in Figure 3 of \citet{Martizzi2018}, the structures that are more likely to be affected by the smoothing length variations are the voids. At redshift $z \gtrsim 4$, the local dark matter density is typically $\sim \Omega_{\rm m} \rho_{\rm crit} > 0.1 \rho_{\rm crit}$, which implies that most of the mass will be assigned to filaments and haloes, rather than voids. At redshift $z \lesssim 4$ the two methods present similar trends for the different cosmic web structures. 
Finally, \citet{Libeskind2018} presents a further analysis of the variations in the mass fractions for each cosmic structure based on different classification methods. As stated in \citet{Martizzi2018}, the differences found between the local dark matter density and the tidal deformation tensor approach are not larger than the deviations presented in \citet{Libeskind2018} for a total of twelve different methods.

To investigate the different gas components, we follow the same classification introduced by \citet{Dave2001} and \citet{Haider2016}, based on the comoving gas density, 
$\rho_{\rm gas}$, and temperature, T, of each gas cell. Four different gas phases are defined in the following way:  
\begin{itemize}
\item \textit{Diffuse gas}, when $\rho_{\rm gas} < 1000 \rho_{\rm crit,0} \Omega_{\rm b}$ and $T<10^{5}$~K. This gas phase is mainly comprised of neutral gas located in the IGM. 
\item \textit{Condensed gas}, when $\rho_{\rm gas} > 1000 \rho_{\rm crit,0} \Omega_{\rm b}$ and $T<10^{5}$~K. 
It selects the gas in star-forming regions and located in the ISM. 
\item \textit{Warm-hot intergalactic medium gas} (WHIM), when $10^{5}~{\rm K} < {\rm T} < 10^{7}$~K. 
This gas phase region represents the circumgalactic medium and hot gas in low-density regions located mainly in the IGM. 
\item \textit{Hot gas}, when ${\rm T} > 10^{7}$~K. This component is mainly composed of shock-heated gas near massive haloes. 
\end{itemize}

As explained by \citet{Martizzi2018}, and \citet{Torrey2019}, IllustrisTNG displays an artificial curvature in the gas phase diagram in the low-density and low-temperature regime, due to a minor numerical heating associated with the expansion of gas in low-density environments. Following the aforementioned papers, we adopt a temperature correction to the gas with density $\rho < 10^{-6} {\rm cm}^3$, in order to ensure that it follows an adiabatic equation of state, ${\rm T }\propto \rho^{\gamma - 1}$. We note, nevertheless, that these corrections do not impact the results of this work in any significant way.

\subsection{Ion abundances}\label{sec:definition_ions}

To compute the fraction of each element in the ionization state of interest, we use the tables developed by \citet{Bird2015}\footnote{See also: \url{https://github.com/sbird/cloudy\_tables}}. The tables were created using {\sc cloudy} version 13.02 \citep{Ferland2013} and calculating both collisional and photo-ionization processes in ionization equilibrium, exposed to a uniform ultraviolet background \citep[UVB][]{Faucher2009}.
{\sc cloudy} was run in single-zone mode adopting constant density and temperature for each gas element. Thus, thermal structures on scales smaller than the gas cells are neglected.
{\sc cloudy} also adopts a frequency dependent self-shielding from UVB at high densities using the fitting function of \citet{Rahmati2013} and solar abundances from \citet{Grevesse2010}. No local radiation sources are considered for running the tables.

The tables computed by \citet{Bird2015} cover a density range of $-7.0 < \log_{10}(n_{\rm H} [{\rm cm}^{-3}]) < 4.0$ and temperatures $3.0 < \log_{10}({\rm T}[{\rm K}]) < 8.6$, assuming a metallicity of ${\rm Z}=0.1{\rm Z_{\odot}}$\footnote{As stated by \citet{Nelson2018_Oxigen} the dependence of metallicity on the fraction of the ionized metals is minor. For this reason and following the same criterion as \citet{Bird2015}, the tables were computed using 1/10 of the solar metallicity.}.
For a given redshift, the ionized fraction of a species in a gas cell is selected from the tabulated values according to its temperature and density. The total mass of a given ionized metal in each gas cell is computed as the ionized fraction selected multiplied by the gas cell mass.

We note that IllustrisTNG adopts a two-phase model for star-forming gas cells \citep[][]{Springel2003}.
This assumption affects the abundance of the species in high-density and low-temperature regions. 
To correct the effect of sub-grid physics, previous studies either place the star-forming cells to ${\rm T}=10^{3}~{\rm K}$ \citep[see e.g.,][]{Nelson2021}, or just simply neglect them from the analysis \citep[][]{DeFelippis2021}. 
In our case, we have decided to keep the star-forming gas cell temperatures provided by the simulation, since they do not generate substantial changes in the results and conclusions of our analysis.

In this work, we study the following species:  C\,\textsc{ii}, C\,\textsc{iv}, Mg\,\textsc{ii}, N\,\textsc{v}, Ne\,\textsc{viii}, O\,\textsc{vi}, and Si\,\textsc{iv}. Each one of these metals maps different regions of the LSS of the Universe. Here we summarize the main properties reported in the literature for each case.
Mg\,\textsc{ii} in emission and absorption is typically detected in star-forming galaxies and planetary nebulae \citep{Rubin2010,Chen2017,Feltre2018}, and has shown to be a good tracer of star-forming regions and cold gas around galaxies \citep[typically at T$\sim10^4$~K, e.g.,][]{Bond2001,Steidel2002,Bouche2007,Weiner2009,Bouche2016,Chen2017}. In addition, different works have associate Mg\,\textsc{ii} absorption systems with galactic haloes \citep[see, e.g.,][]{Lanzetta1990}.  Mg\,\textsc{ii} has a ionization potential of $15.03~{\rm eV}$.
C\,\textsc{ii} is a good tracer of dense and low-temperature gas (n$_{\rm H} = 10^{2} - 10^{3}{\rm cm}^{-3}$, ${\rm T\sim 10^{4}} {\rm K}$ and ionization potential of $24.38~{\rm eV}$), associated with galactic haloes as well. It is also a star formation rate tracer and a major coolant for neutral atomic gas in the ISM \citep{Stacey1991,Wolfire2003,Stacey2010,Herrera-Camus2015}. C\,\textsc{iv} is a tracer of enriched gas in the IGM (with ${\rm n_{H} }\sim 10^{-1} {\rm cm^{-3}}$ and ionization potential of $64.49~{\rm eV}$). It is also mentioned as a good tracer of hot gas located in shock-heated regions and CGM \citep{DOdorico2010,Cooksey2010,
Danforth2016,Bouche2016,Chen2017}. Si\,\textsc{iv} in absorption is a strong doublet of Si and a good tracer of $\alpha$-elements and young metals. Si\,\textsc{iv} has a ionization potential of $45.14~{\rm eV}$, and is also typically present in high-density gas clouds \citep{Cooksey2011}.
Ne\,\textsc{viii} is mainly found in regions of gas at low densities and high temperatures  \citep[T~$\sim 10^5 - 10^6$~K, and n$_{H}\gtrsim10^{-4}~{\rm K}$, e.g.,][]{Meiring2013,Burchett2019}. Ne\,\textsc{viii} has a ionization potential of $239.09~{\rm eV}$.
N\,\textsc{v} has been shown to be a good tracer of the warm-hot CGM \citep[][]{Danforth2008}, and has a ionization potential of $97.89~{\rm eV}$.
O\,\textsc{vi} is a well-known tracer of warm hot gas in low-dense regions, typically located in galactic haloes and the CGM \citep[at T~$\gtrsim 2\times 10^4$~K, e.g.,][]{Werk2014,Roca-Fabrega2019}, with a ionization potential of $138.12~{\rm eV}$.
These features represent only a partial summary of each of the ionized metals studied in this work. We refer to the aforementioned references for further details.

\section{Results}\label{sec:results}

The results presented in this work address two main 
aspects that are relevant to the missing baryons problem. First, we discuss the baryonic mass fraction (gas and stars) within dark matter haloes, as an extension of the results presented by \citet{Haider2016} for Illustris and \citet{Martizzi2018} for IllustrisTNG (Section~\ref{sec:baryons_haloes}).
The second part explores the distribution of ionized metals in the cosmic web (Section~\ref{sec:ions_cosmicWeb}) and in the density-temperature phase space (Section~\ref{sec:ions_phases}) across cosmic time.

\subsection{Baryon mass fraction in haloes}
\label{sec:baryons_haloes}

%%%%%%%%%%%%%%%%%%%%%%%%%%%
%%%%%%%%%%%%%%%%%%%%%%%%%%%
%%%%%%%%%  FIG 2  %%%%%%%%%%
\begin{figure*}
\centering
%/n/home01/mcartale/MARKUS_FOLDERS/illustrisTNG/baryonFractionInHalos
%This plot is done in the cluster 
\includegraphics[width=0.8\textwidth]{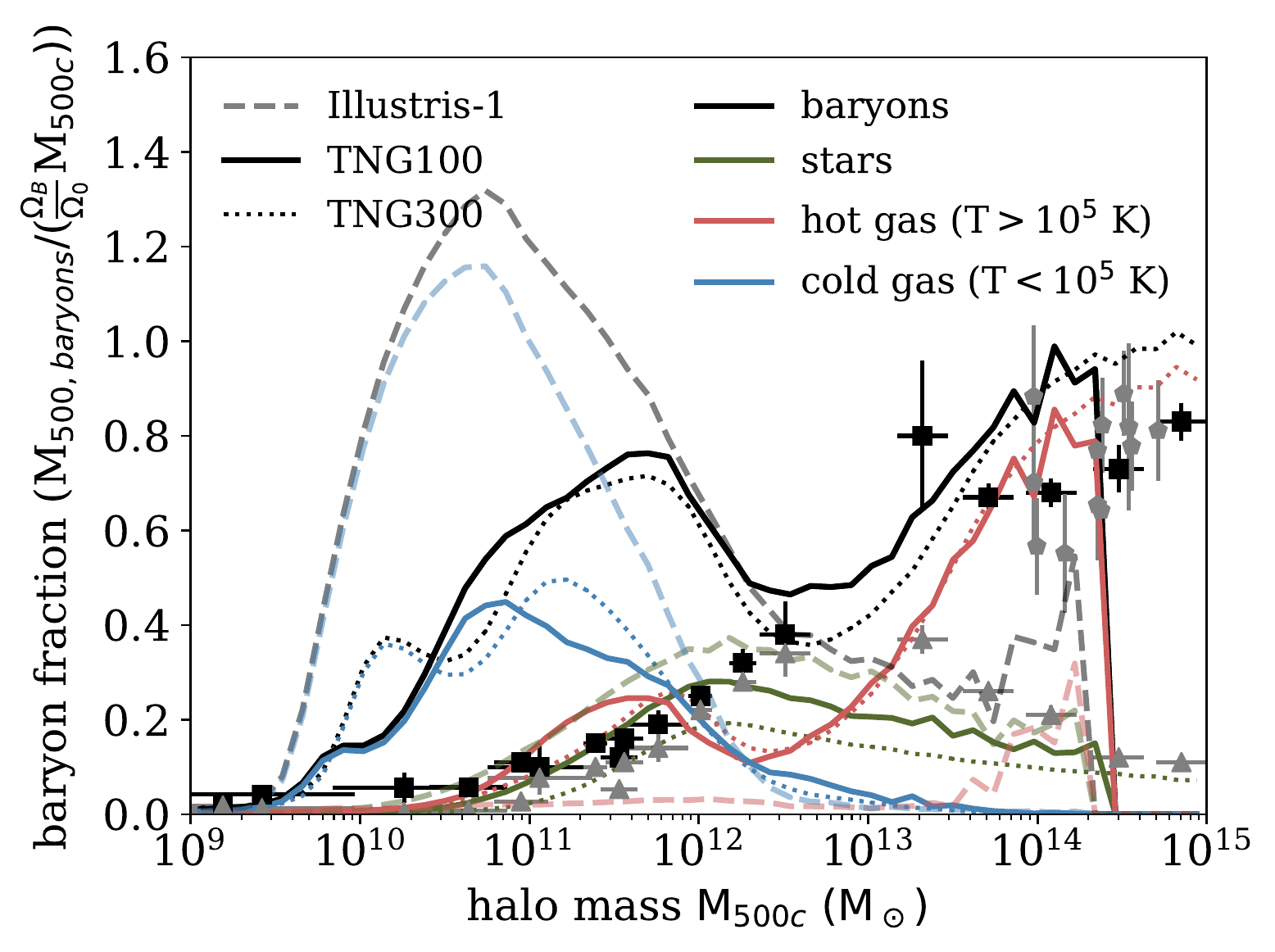}
\caption{Baryon fraction as a function of the dark matter halo mass (M$_{500c}$). We compute the baryon and dark matter mass 
within R$_{500c}$ at $z=0$ for Illustris (dashed lines), TNG100 (solid lines), and TNG300 (dotted lines).
We compare our results with observational data from \citet{McGaugh2010} (baryons: black squares, stars: grey triangles) and \citet{Gonzalez2013} (grey diamonds). The observational results from \citet{McGaugh2010} belongs to a set of galaxies (from dwarf to clusters) for which the baryonic and stellar fractions were computed according to the gas and stellar dynamics. As stated by \citet{McGaugh2010}, a significant gas fraction might be hidden in a state unable to be detected, suggesting that the reported fraction is incomplete. The observational data from \citet{Gonzalez2013} corresponds to the hot-gas from twelve galaxy clusters.}
\label{fig:baryonsInHalos}
\end{figure*}
%%%%%%%%%%%%%%%%%%%%%%%%%%%
%%%%%%%%%%%%%%%%%%%%%%%%%%%
%%%%%%%%%%%%%%%%%%%%%%%%%%%

The baryon mass fraction is computed as the total baryonic mass within the radius at which the density is equal to 500 times the critical density, M$_{\rm 500,baryons}$, normalized by the relative cosmological value, i.e., $[\Omega_{\rm b}/\Omega_{m}]{\rm M}_{\rm 500c}$. %This is typically the common approach to estimate the deviations with respect to the cosmic fraction in haloes. We note, however, that these deviations might vary based on the halo definition adopted.
Figure~\ref{fig:baryonsInHalos} presents the baryon mass fraction as a function of the halo mass at $z=0$, for haloes with M$_{\rm 500c} > 10^{9} {\rm M}_{\odot}$.
To analyse the impact of resolution and subgrid models, Figure~\ref{fig:baryonsInHalos} displays the results obtained for Illustris, TNG100, and TNG300.
We split the contribution from stars, hot gas (${\rm T}>10^{5}$K), and cold gas (${\rm T}<10^{5}$K) and compare them with the observational data from \citet{McGaugh2010} and \citet{Gonzalez2013}. Our results show that the baryonic mass fractions of TNG100 and TNG300 follow similar trends for almost the entire halo mass range under analysis, with differences mostly caused by resolution/volume effects.
Namely, TNG300 is able to reproduce the observations for massive clusters (${\rm M}_{500c} > 10^{13} {\rm M}_{\odot}$), reaching values between 0.75 to 1 for the baryon fraction (gas and stars) and 0.1 to 0.3 for the stellar fraction. The TNG100 box reproduces well the observations for clusters with masses above $10^{13} {\rm M}_{\odot}$ and up to  $2 \times 10^{14} {\rm M}_{\odot}$. The very massive end is not accessible in TNG100 due to its small size, which explains the artificial drop in the baryonic fraction shown for the largest clusters in Figure~\ref{fig:baryonsInHalos}. As shown also in \citet{Genel2014} and \citet{Haider2016}, the Illustris simulation is unable to reproduce the observed total baryon fraction in massive clusters above ${\rm M}_{500c} \simeq 10^{13} {\rm M}_{\odot}$.

For haloes with masses ${\rm M}_{500c} < 10^{13} {\rm M}_{\odot}$, the stellar mass fractions of TNG100 and TNG300 are in reasonable agreement with the observational constraints reported by \citet{McGaugh2010}. The missing baryon problem from observational data manifests itself in the vanishing 
baryonic mass fraction below ${\rm M}_{500c} \simeq 10^{12} {\rm M}_{\odot}$. The simulated haloes from the three boxes display, conversely, a high baryon mass fraction for low-mass haloes, reaching six times the observed fraction in the TNG300 box and twenty times in the Illustris simulation at $\sim 5\times 10^{10}{\rm M}_{\odot}$, respectively. It is noteworthy, however, that observational estimates might be severely affected by uncertainties: current observations are not able to detect the different gas phases within haloes in the way we do in simulations. In fact, more recent measurements of the baryonic mass fraction for haloes with masses around $10^{12}{\rm M}_{\odot}$ from the COS-Halos survey claim a significantly higher value (45\%), suggesting as well that at least half of the previously reported missing baryons might be located in the CGM \citep{Werk2014}. The comparison presented in Figure~\ref{fig:baryonsInHalos} must therefore be taken with caution, since it is subject to uncertainties in the observational constraints that are not necessarily accounted for by the reported errors. 

The general agreement between TNG100 and TNG300 in Figure~\ref{fig:baryonsInHalos} is expected since both simulation boxes use the same sub-grid model. 
This is not the case for Illustris, which explains the differences with respect to the aforementioned IllustrisTNG boxes. 
In particular, previous reports have identified 
several caveats in the Illustris stellar and AGN feedback models \citep{Genel2014,Haider2016, Pillepich2018}. The consequences of these issues are actually noticeable in Figure~\ref{fig:baryonsInHalos}: the stellar feedback model in Illustris is not efficient in removing gas in dark matter haloes below $10^{12} {\rm M}_{\odot}$, while the radio-mode AGN feedback produces strong outflows mainly in haloes above 10$^{12.5}$~M$_{\odot}$.
These processes have been revised in the updated IllustrisTNG model, where AGN feedback is now described in terms of low and high accretion states that account for kinetic and thermal feedback.
For haloes with ${\rm M}_{500c} < 10^{12} {\rm M}_{\odot}$, supernova feedback prevents star formation and removes gas through an updated model for galactic winds.

\begin{figure}
\centering
%Computed in:/n/home01/mcartale/MARKUS_FOLDERS/illustrisTNG/halo_evolution/
%This plot is in /home/celeste/Dropbox/IllustrisTNG/EvolutionOfMass
% In Fraction_In_Halos_vCel.py
\includegraphics[width=0.5\textwidth]{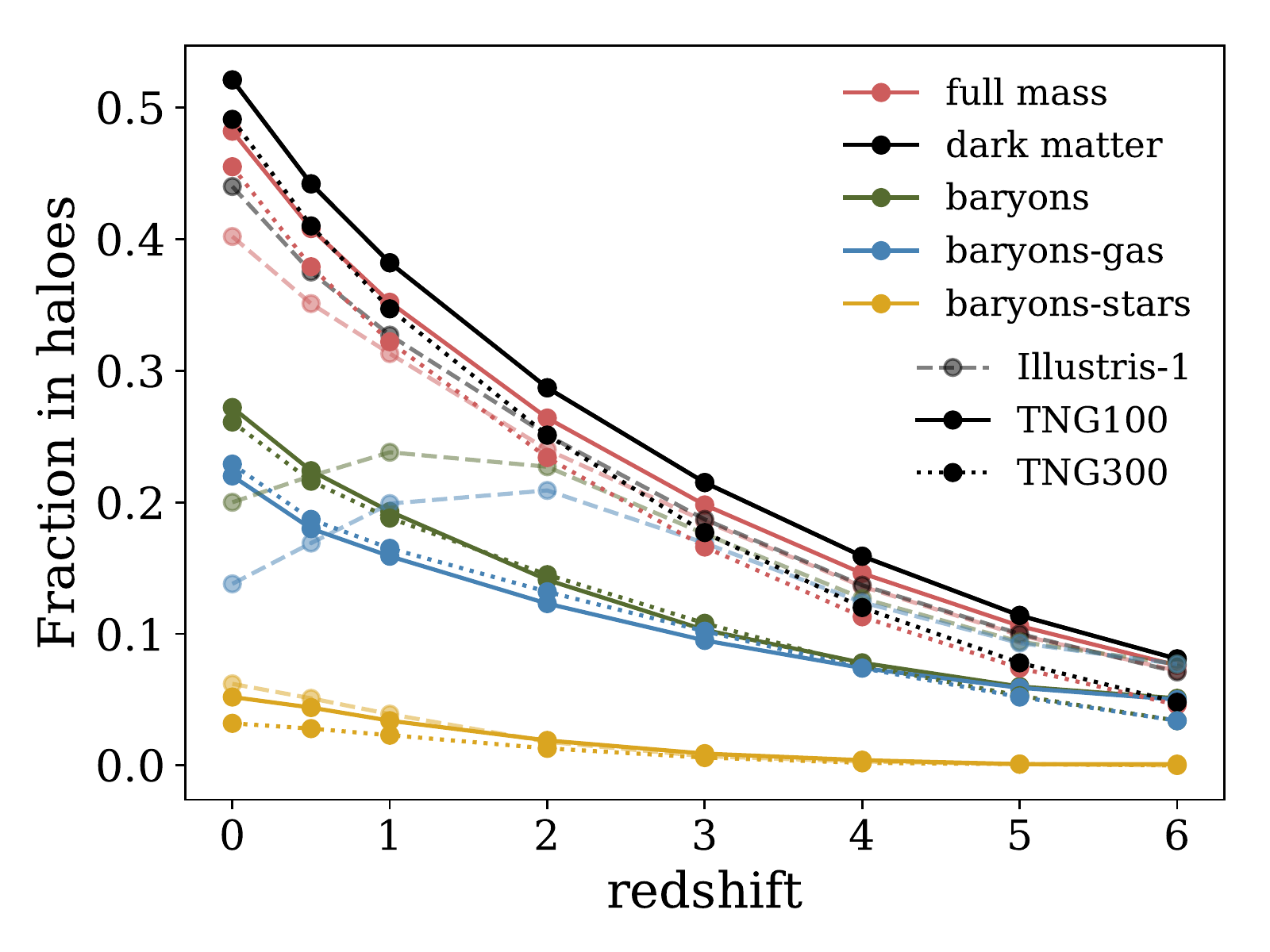}%halo_evo_205.pdf}
\caption{Evolution of the mass fraction within dark matter haloes for TNG100 (solid lines), TNG300 (dotted lines) and Illustris-1 (dashed lines).
We show the mass fraction of dark matter (black lines), baryons (green lines), and the contribution from gas (cyan lines) and stars (yellow lines).
Each of these fractions is computed by normalizing by its respective total mass in each simulated volume.}
\label{fig:halo_evolution}
\end{figure}

We proceed now to address the redshift evolution 
of the mass fractions. These results are shown for haloes in Illustris, TNG100, and TNG300 in 
Figure~\ref{fig:halo_evolution}. The mass fraction is computed for dark matter, baryons (split also in gas and stars), and the full mass (baryons and dark matter), normalized by its respective total mass for each case. We find that dark matter mass in haloes increases with time for the three boxes, with values that range from $\sim 5-7$\% at $z=6$ to $\sim 40-55$\% at $z=0$ . The same behaviour is, as expected, found for the full mass fraction, since dark matter dominates the mass budget of haloes. 

The baryon fraction in haloes shows a different behaviour in Illustris as compared to IllustrisTNG at $z < 2$. The difference lies in the gas fraction: while TNG100 and TNG300 show a monotonic increase as the redshift decreases, Illustris presents an artificial drop in the gas fraction. As stated by \citet{Haider2016}, the decrease in the gas fraction is connected with the time that the radio mode AGN feedback becomes overly efficient at expelling gas from the inner regions to the outskirts. As Figure~\ref{fig:halo_evolution} demonstrates, the updated AGN feedback model in IllustrisTNG alleviates this problem. 

The results presented in this section demonstrate that IllustrisTNG provides better agreement with observations than Illustris, which is a consequence of the improved subgrid model of the former. In the next section, we will go a step further and investigate the large-scale distribution of ionized metals.

%%%%%%%%%%%%%%%%%%%%%%%%%%%
%%%%%%%%%%%%%%%%%%%%%%%%%%%
%%%%%%%%%  FIG 3  %%%%%%%%%%
%%%%%%%%%%%%%%%%%%%%%%%%%%%
%%%%%%%%%%%%%%%%%%%%%%%%%%%
%%%%%%%%%%%%%%%%%%%%%%%%%%%

\subsection{Ionized metals in the Cosmic Web}
\label{sec:ions_cosmicWeb}

%%%%%%%%%%%%%%%%%%%%%%%%%%%
%%%%%%%%%%%%%%%%%%%%%%%%%%%
%%%%%%%%%  FIG 4  %%%%%%%%%%
\begin{figure*}
\centering
\includegraphics[width=0.47\textwidth]{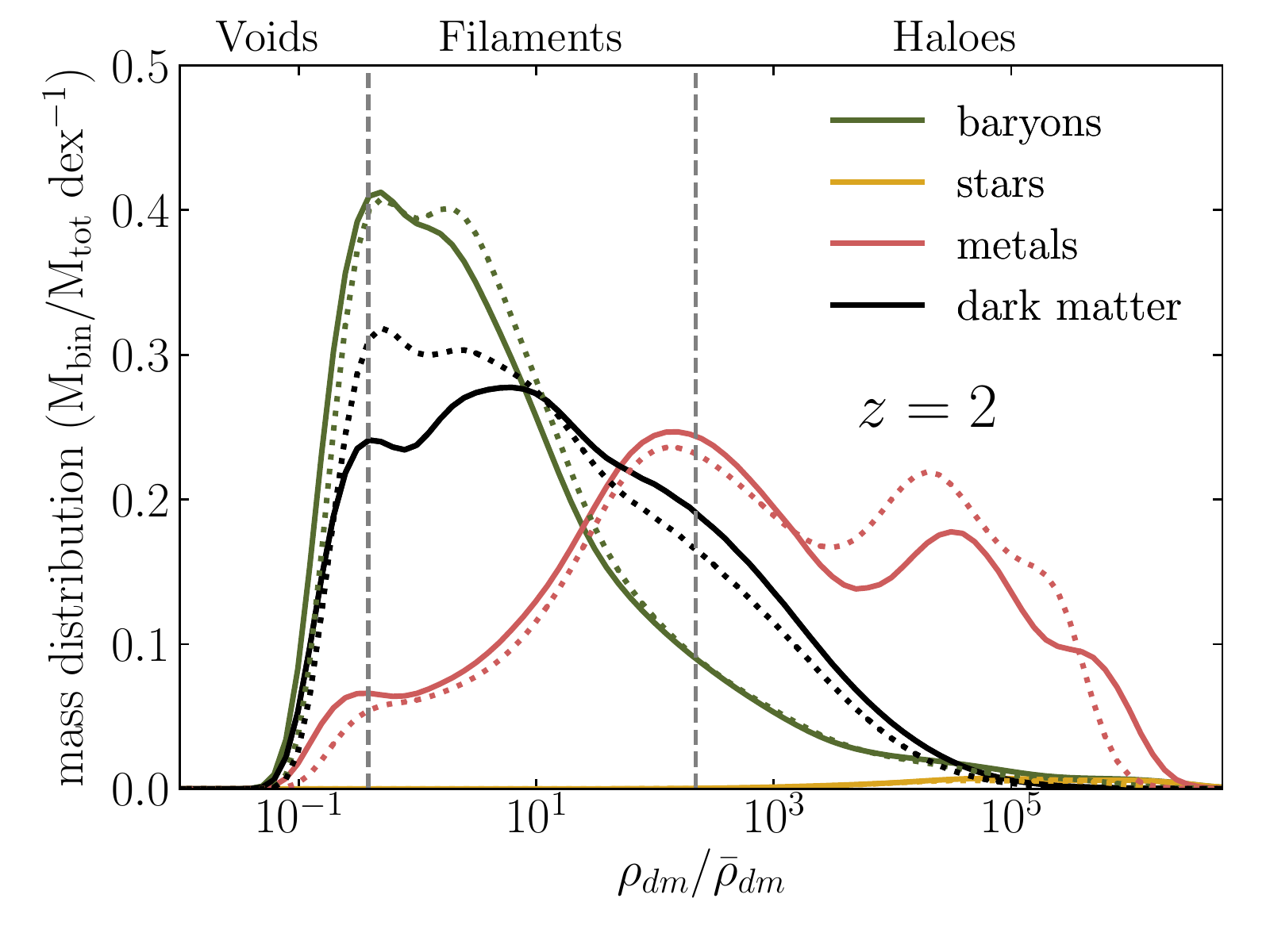}
\includegraphics[width=0.47\textwidth]{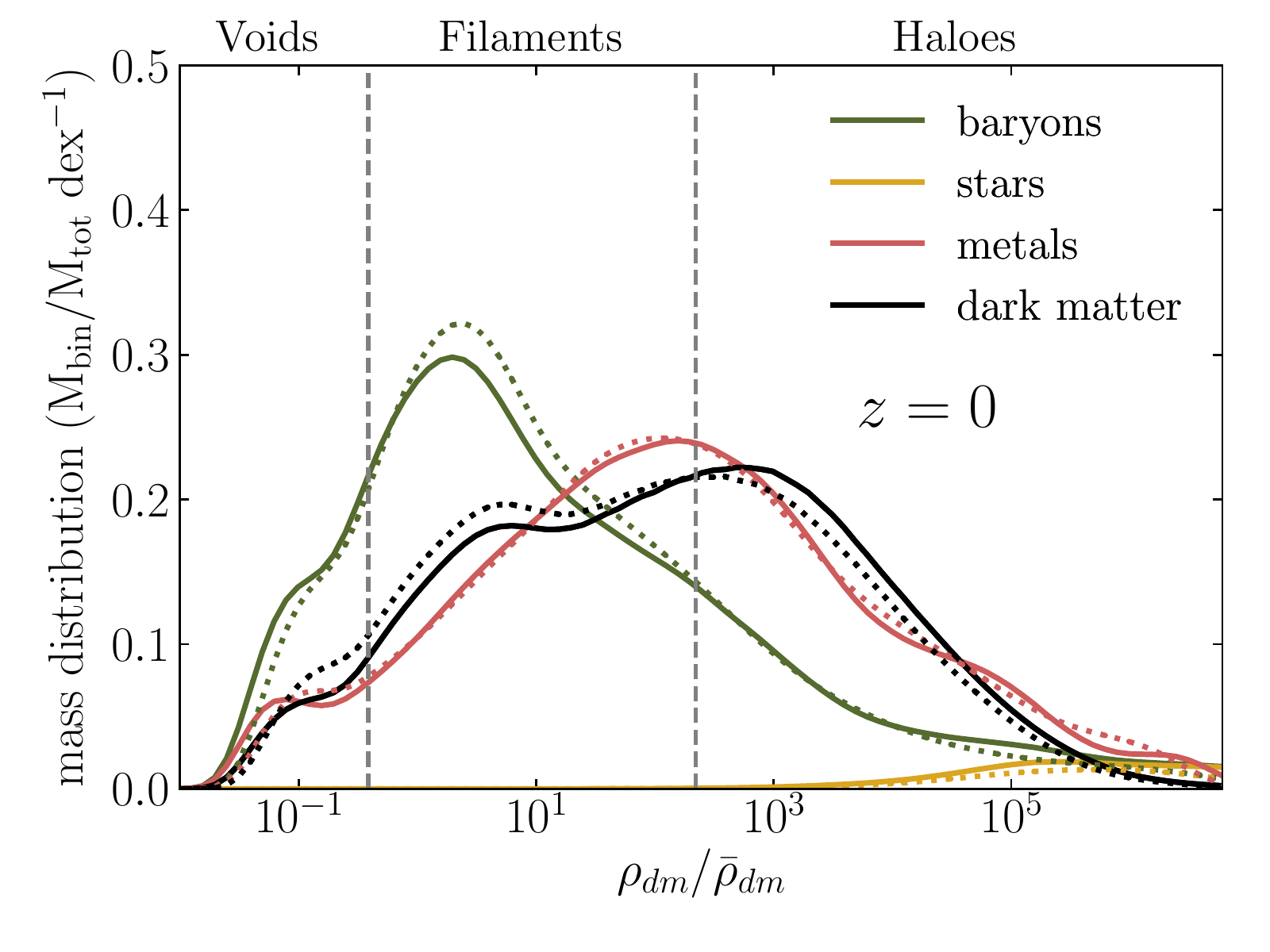}
\caption{Mass distribution as a function of the local dark matter density, normalized by the mean density for TNG100 (solid lines), and TNG300 (dotted lines) at $z=0$ and 2.  
The dashed vertical lines indicate the boundaries between the voids, filaments and haloes.}
\label{fig:mass_distribution}
 \end{figure*}
%%%%%%%%%%%%%%%%%%%%%%%%%%%
%%%%%%%%%%%%%%%%%%%%%%%%%%%
%%%%%%%%%%%%%%%%%%%%%%%%%%%

%%%%%%%%%%%%%%%%%%%%%%%%%%%
%%%%%%%%%%%%%%%%%%%%%%%%%%%
%%%%%%%%%  FIG 5  %%%%%%%%%%
\begin{figure*}
\centering
\includegraphics[width=0.47\textwidth]{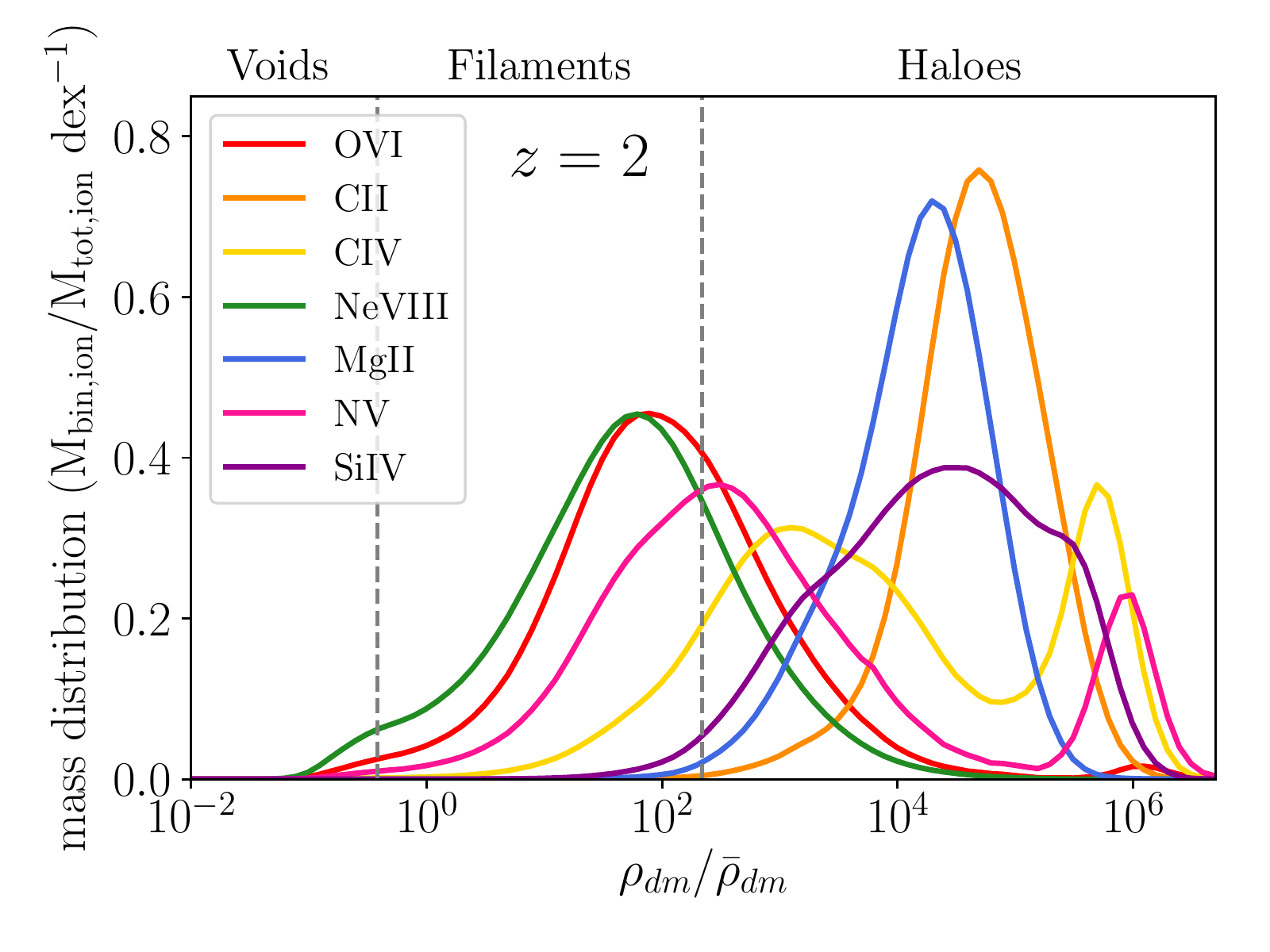}
\includegraphics[width=0.45\textwidth]{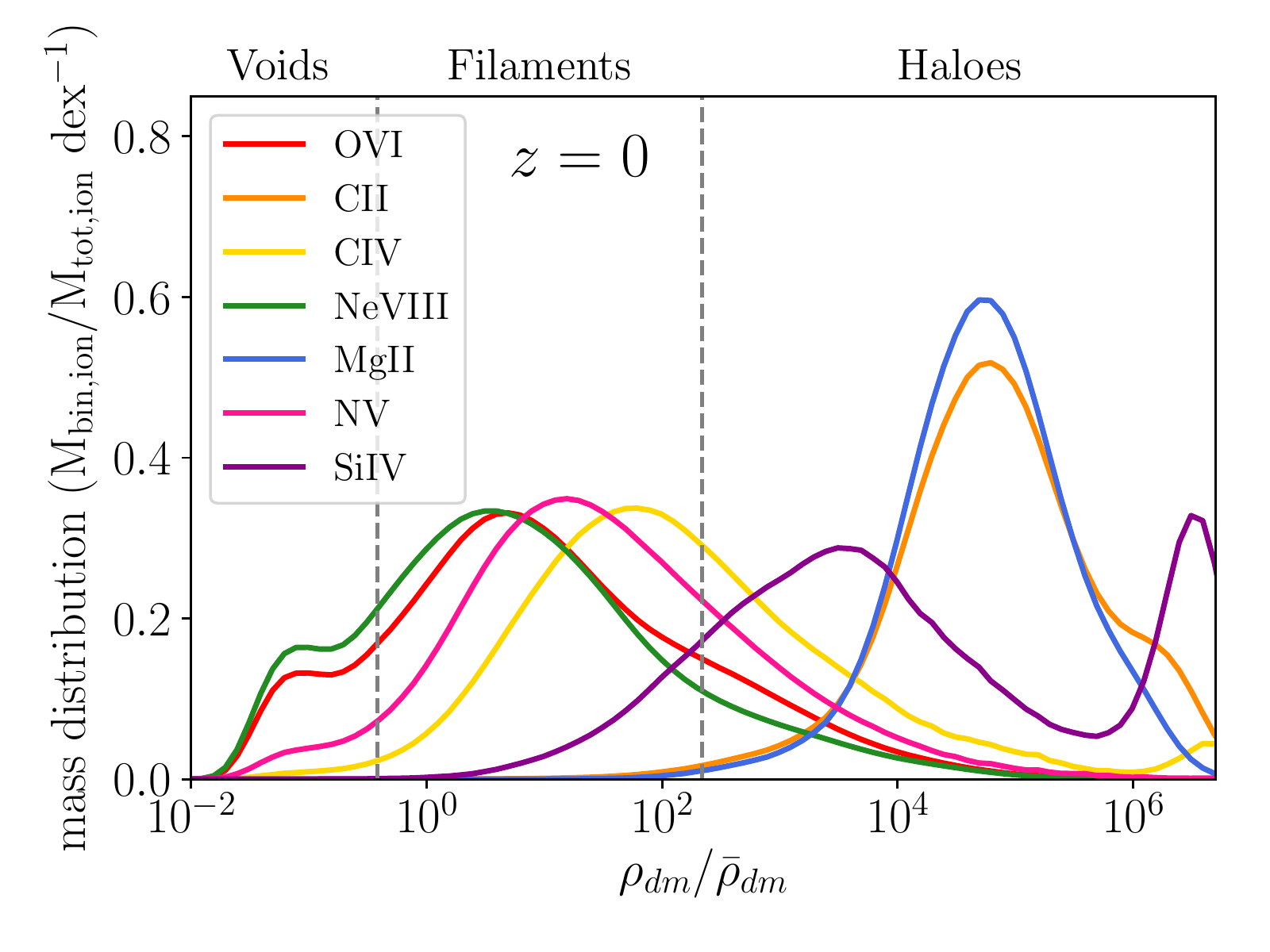}
\caption{Fraction of ions, as a function of the local dark matter density at $z=2$ (\textit{left panel}) and $z=0$ (\textit{right panel}),
for TNG100. The fraction of ions is computed as the mass at a fixed local dark matter density, normalized by the total mass of each ion.
The local dark matter density is also normalized by the mean dark matter of the simulated box.
Vertical black dotted lines represent the criterion adopted to identify the haloes, filaments and voids. The double peak of C\,\textsc{iv} and N\,\textsc{v} at $z=2$, and of Si\,\textsc{iv} at $z=0$ must be interpreted with caution, since these peaks are produced by star-forming gas affected by the sub-grid physics of the simulation (see Section~\ref{sec:definition_ions}).}
\label{fig:ions_distribution}
 \end{figure*}
%%%%%%%%%%%%%%%%%%%%%%%%%%%
%%%%%%%%%%%%%%%%%%%%%%%%%%%
%%%%%%%%%%%%%%%%%%%%%%%%%%%

We begin by analyzing the performance of the method adopted to trace the cosmic web (described in Section~\ref{sec:definition_cosmicWeb}). Figure~\ref{fig:mass_distribution} shows the distribution of the mass fraction of baryons, stars, metals\footnote{The mass fraction of metals (which in turn represents a portion of the baryons) is computed by selecting all the metal elements heavier than He.}, and dark matter at $z=0$ and 2 
in TNG100 and TNG300, as a function of the comoving local dark matter density normalized by its mean density (i.e., $\rho_{dm}/\bar{\rho}_{dm}$, where $\bar{\rho}_{dm} = \Omega_{dm}$ \rhocrit). Dotted vertical lines indicate the criterion to define voids ($\rho_{dm} < 0.1$~\rhocrit), filaments ($\rho_{dm}$ = 0.1 - 57~\rhocrit), and haloes ($\rho_{dm} > 57$~\rhocrit).

From the comparison of the two panels of Figure~\ref{fig:mass_distribution}, we find that metals slightly move from high-density to low-density regions as the redshift decreases. This is probably a consequence of the outflows and chemical enrichment produced by the stellar and AGN feedback processes, which tend to transport enriched gas from haloes to voids and filaments. 
Conversely, dark matter is unaffected by baryonic processes on large scales. It continues assembling and condensing in filaments and haloes, as Figure~\ref{fig:mass_distribution} demonstrates. Good agreement is again found between the distributions obtained with TNG100 and TNG300, confirming that resolution does not affect the results and main conclusions of this work in any significant way.
For simplicity, the analysis of the ionized metals is presented in what follows only for the higher-resolution TNG100 box, since similar trends are obtained for the TNG300 box.
Finally, the cosmic evolution of the total baryonic budget in IllustrisTNG is discussed in detail in \citet{Martizzi2018,Martizzi2020} and is out of the scope of this work. We review some of the main features in Appendix~\ref{fig:Appendix_CosmicStructures}.

Figure~\ref{fig:ions_distribution} displays the distribution of our set of ionized metals as a function of the local dark matter density normalized by its mean density, $\rho_{dm}/\bar{\rho}_{dm}$, at $z=0$ and 2 for TNG100. The fraction of ions is computed as the mass of each ion normalized by its total mass at a fixed local dark matter density. Again, dotted vertical lines represent the criterion to define voids, filaments and haloes. Our results show that C\,\textsc{ii}, Mg\,\textsc{ii} and Si\,\textsc{iv} are significantly more abundant in haloes than in filaments and voids, both at $z=2$ and 0. At $z=2$ these three ions concentrate in haloes with local dark matter densities around $\rho_{dm}/\bar{\rho}_{dm} \sim 10^4 - 10^5$. At $z=0$, the distribution remains in the same range for Mg\,\textsc{ii} and C\,\textsc{ii} but splits into two peaks at low and very high densities for Si\,\textsc{iv}. O\,\textsc{vi}, C\,\textsc{iv}, Ne\,\textsc{viii}, and N\,\textsc{v} tend to move from haloes and filaments at $z=2$ to voids and filaments at $z=0$ (towards lower-density regions in general). These ions display a peak in their fractions for local dark matter densities of $\rho_{dm}/\bar{\rho}_{dm} \sim 10 - 10^2$ at $z=0$.  Figure~\ref{fig:ions_distribution} also shows a bimodal distribution for C\,\textsc{iv} and N\,\textsc{v} at $z=2$, reflecting the presence of collisionally ionized gas in dense regions. The same result is found for Si\,\textsc{iv} at $z=0$. These features, however, must be interpreted with caution, since they are produced by ionized metals in star-forming gas cells (see Section~\ref{sec:definition_ions}). In order to investigate the impact of the sub-grid model on the mass distributions, we have replaced 
the tabulated temperature of the star-forming cells by a fixed value of 10$^3$~K \citep[based on other recent analyses,][]{Nelson2021}. We have checked that imposing this value effectively removes the double peak in C\,\textsc{iv}, N\,\textsc{v}, and Si\,\textsc{iv}, while maintaining the overall shape of the mass distributions displayed in Figure~\ref{fig:ions_distribution}.

%%%%%%%%%%%%%%%%%%%%%%%%%%%
%%%%%%%%%%%%%%%%%%%%%%%%%%%
%%%%%%%%%  FIG 6  %%%%%%%%%%
\begin{figure*}
\centering
\includegraphics[width=0.95\textwidth]{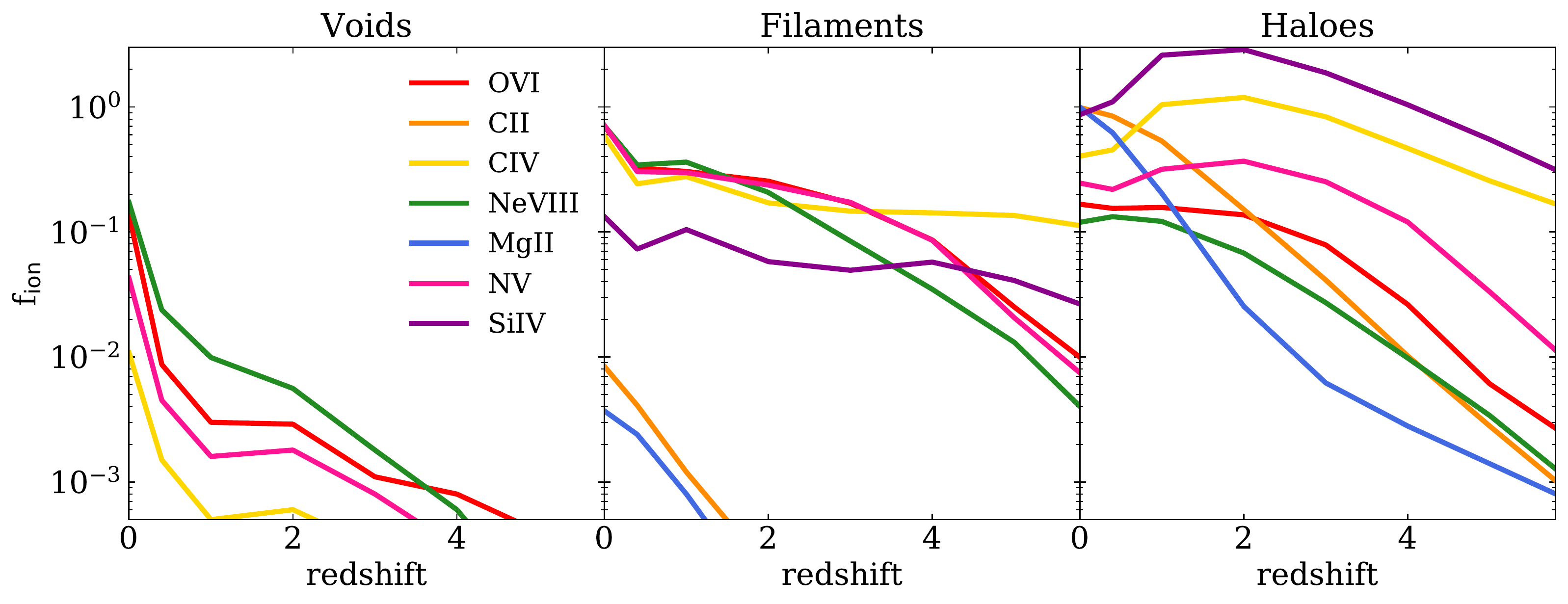}
\caption{Fraction of the ionized metals as a function of redshift, split in voids, filaments and haloes for TNG100. 
It is computed as the total mass of each ionized metal at a given redshift
normalized by its total mass budget at $z=0$, (i.e., $f_{\rm ion} = M_{X, z} / M_{X,z=0}$). 
}
\label{fig:ions_VoidsFilamentsHaloes}
\end{figure*}
%%%%%%%%%%%%%%%%%%%%%%%%%%%
%%%%%%%%%%%%%%%%%%%%%%%%%%%
%%%%%%%%%%%%%%%%%%%%%%%%%%%

To investigate in more detail the evolution of each ion, Figure~\ref{fig:ions_VoidsFilamentsHaloes} shows the mass fraction for each ionized metal in voids, filaments, and haloes normalized by the total mass of each ion at $z=0$ (i.e., f$_{\rm ion} = M_{X, z} / M_{X,z=0}$) for the redshifts $z=0-6$. Using f$_{\rm ion}$ allows us to trace the evolution of the mass budget with redshift in an effective way since the abundance of metals increases with time.
At first glance, our results show that most of the mass budget for all the ionized metals under analysis is either in filaments or haloes throughout the entire redshift range considered. However, each one presents different properties and evolution across cosmic time. We now proceed to analyse the contribution of the ions to the mass budget in each cosmic web region. In voids, the highest contribution belongs to O\,\textsc{vi} and Ne\,\textsc{viii}, but 
N\,\textsc{v} and C\,\textsc{iv} are also present. All of these ions increase their mass fractions towards lower redshift. C\,\textsc{ii}, Mg\,\textsc{ii}, and Si\,\textsc{iv} are not found in this cosmic web environment.

For filaments, O\,\textsc{vi}, N\,\textsc{v}, and Ne\,\textsc{viii} show a similar and significant increase of their mass fractions as redshift decrease, of a factor $\sim$60 from $z=6$ to $z=0$. We note that most of the mass budget for Ne\,\textsc{viii} and O\,\textsc{vi} is located in this region of the cosmic web. Their large abundance in filaments has been reported by previous observational and numerical studies, showing that O\,\textsc{vi} is an excellent tracer of filamentary diffuse and warm gas between clusters \citep[see e.g.,][]{Nicastro2017,Nicastro2018,Tepper-Garcia2011,Tejos2016,Marra2020,Strawn2020}. This characteristic makes O\,\textsc{vi} arguably one of the best ``footprints" of missing baryons together with O\,\textsc{vii} and O\,\textsc{viii}. The evolution of the mass fractions for O\,\textsc{vi}, N\,\textsc{v}, and Ne\,\textsc{viii}
is determined by the chemical enrichment of the gas due to stellar evolution and feedback, together with virial shock-heating increasing its temperature \citep[see e.g.,][]{Tepper-Garcia2011,Faerman2017}. It is also noteworthy that the mass fractions of C\,\textsc{iv} and Si\,\textsc{iv} in filaments are roughly flat, showing almost no redshift dependence.  
Finally, the contribution of C\,\textsc{ii} and Mg\,\textsc{ii} is only visible, although tiny ($<1\%$), at low redshift, around $z\sim0-2$.

Figure~\ref{fig:ions_VoidsFilamentsHaloes} shows that haloes contain all ionized metals over the entire redshift range, albeit in different proportions.  Interestingly, two different behaviours are observed for the set of ions investigated. While all metals begin increasing their fractions at high redshift, Si\,\textsc{iv}, C\,\textsc{iv}, N\,\textsc{v}, and O\,\textsc{vi}, seems to reach their peak abundances around $z\sim2$, which roughly coincides with the peak of the cosmic star formation rate. At lower redshifts, the evolution of these ions either stalls or their abundances actually drop. Conversely, the fractions of C\,\textsc{ii} and Mg\,\textsc{ii} continue rising towards lower redshift and only seem to reach a plateau, if anything, at $z=0$. At face value, these results indicate that Si\,\textsc{iv}, C\,\textsc{iv}, N\,\textsc{v}, and O\,\textsc{vi} in haloes are strongly affected by stellar and AGN feedback mechanisms, together with gas recycling and radiative cooling. While the total mass budget of each metal increases with time, the ionization species are influenced by gas cooling, and star formation in dense regions. The contribution of all these effects together may explain the cosmic evolution of the mass fractions for Si\,\textsc{iv}, C\,\textsc{iv}, N\,\textsc{v}, and O\,\textsc{vi} in haloes.

\subsection{Ionized metals in density-temperature phase space}
\label{sec:ions_phases}

%%%%%%%%%%%%%%%%%%%%%%%%%%%
%%%%%%%%%%%%%%%%%%%%%%%%%%%
%%%%%%%%%  FIG 7  %%%%%%%%%%

\begin{figure*}
\centering
\includegraphics[width=1.\textwidth]{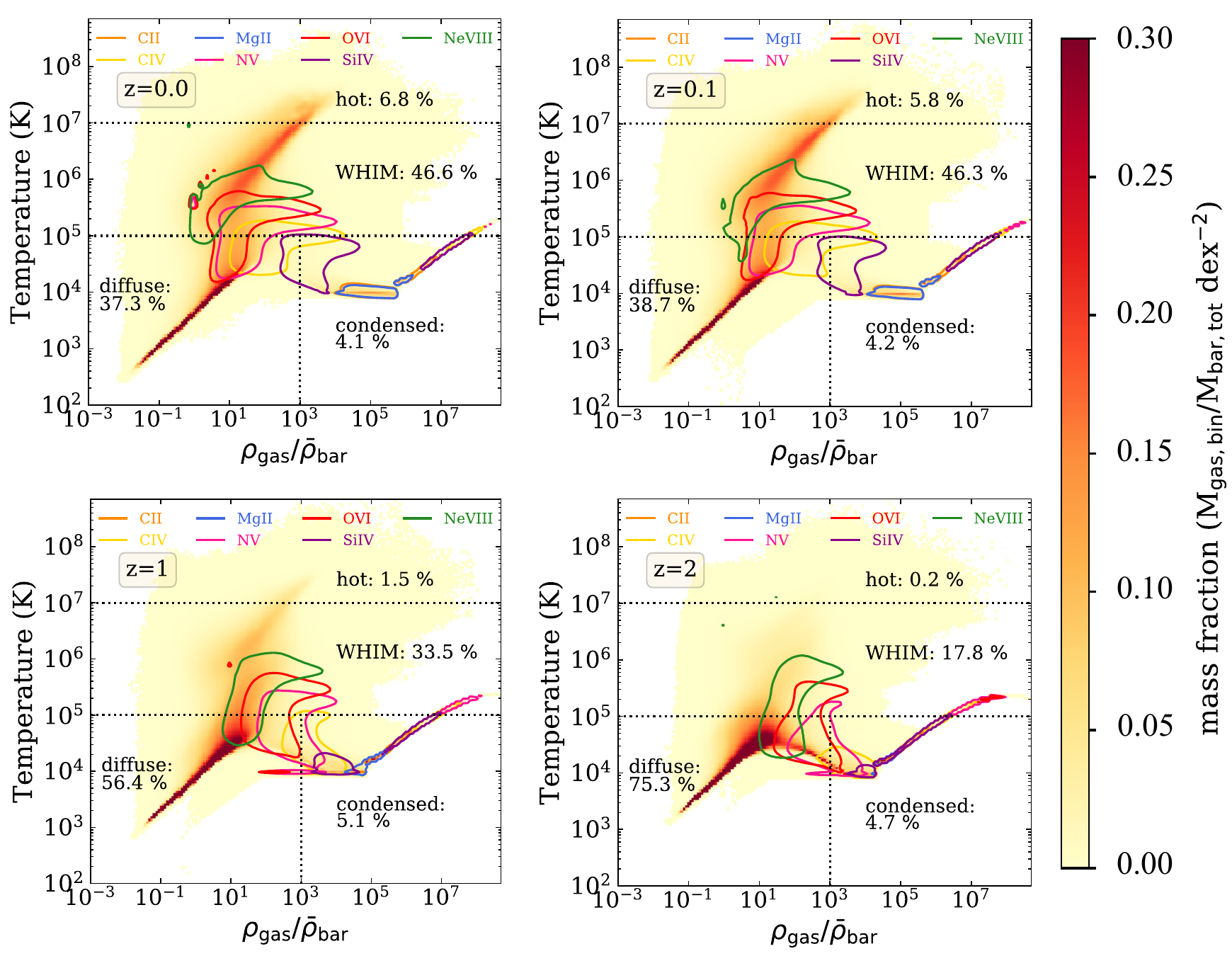}
\caption{Gas density--temperature diagram at redshifts $z=0, 0.1, 1$, and 2 (from top left to bottom right, respectively) for the TNG100 simulation.
The colour map represents the mass fraction of gas, while the contours indicate the regions enclosing 80 per-cent the mass of the ionized metals Mg\,\textsc{ii}, Ne\,\textsc{viii}, N\,\textsc{v}, O\,\textsc{vi}, Si\,\textsc{iv}, C\,\textsc{ii} and C\,\textsc{iv}. 
Black dashed lines demarcate the gas phases adopted in this work (hot, WHIM, diffuse, and condensed). }
\label{fig:T_rho_ions}
\end{figure*}

%%%%%%%%%%%%%%%%%%%%%%%%%%%
%%%%%%%%%%%%%%%%%%%%%%%%%%%
%%%%%%%%%  FIG 8  %%%%%%%%%%
\begin{figure*}
\centering
\begin{subfigure}{0.24\linewidth}
 \caption{\centering Voids+Filaments+Haloes}
  \includegraphics[width=\textwidth]{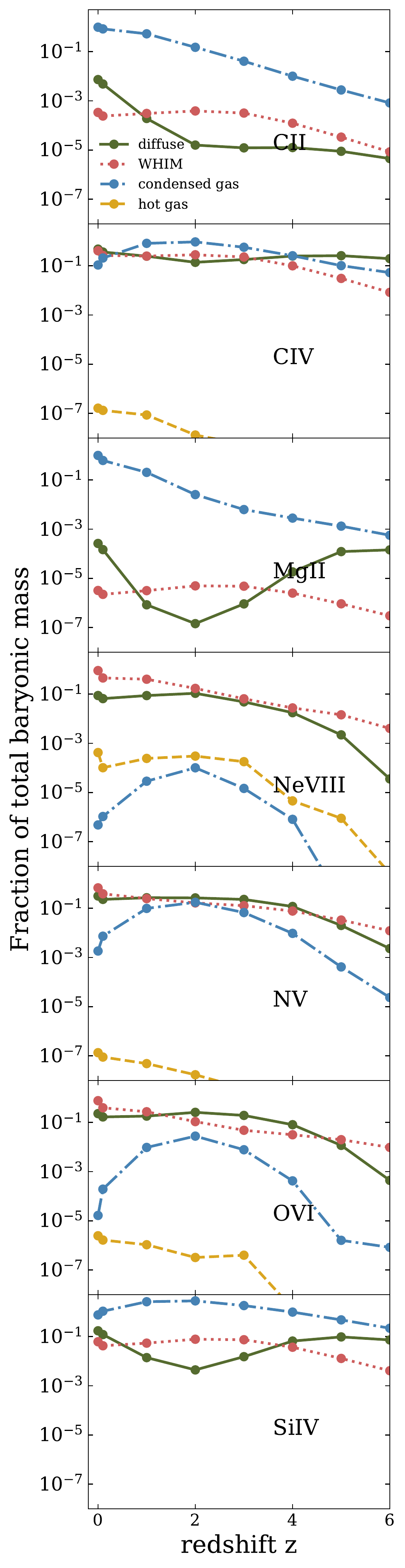}
 \end{subfigure}
\begin{subfigure}{0.24\linewidth}
 \caption{\centering Voids}
 \includegraphics[width=\textwidth]{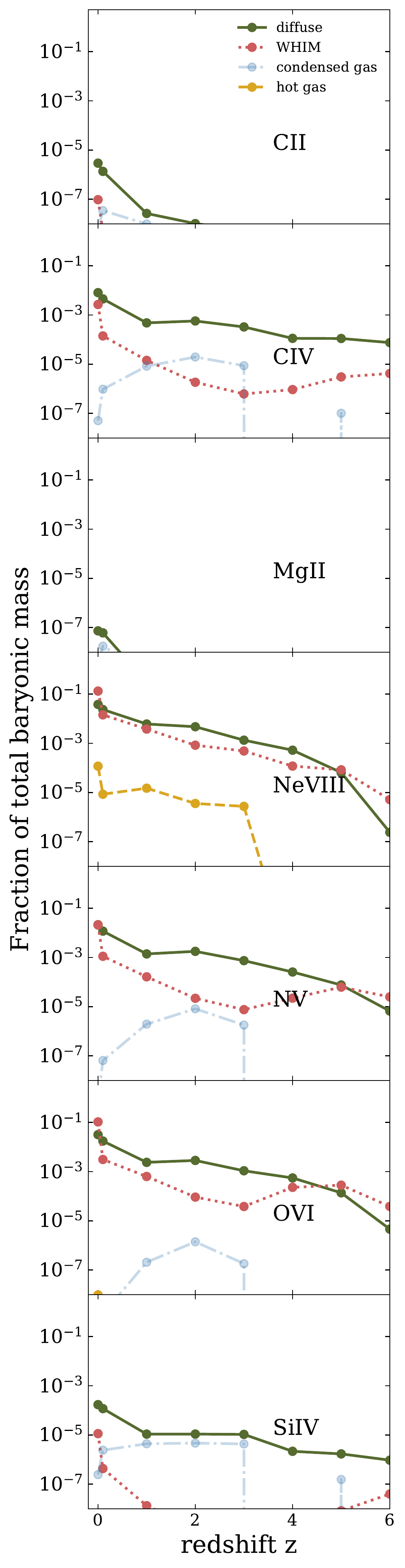}
 \end{subfigure}
 \begin{subfigure}{0.24\linewidth}
 \caption{\centering Filaments}
 \includegraphics[width=\textwidth]{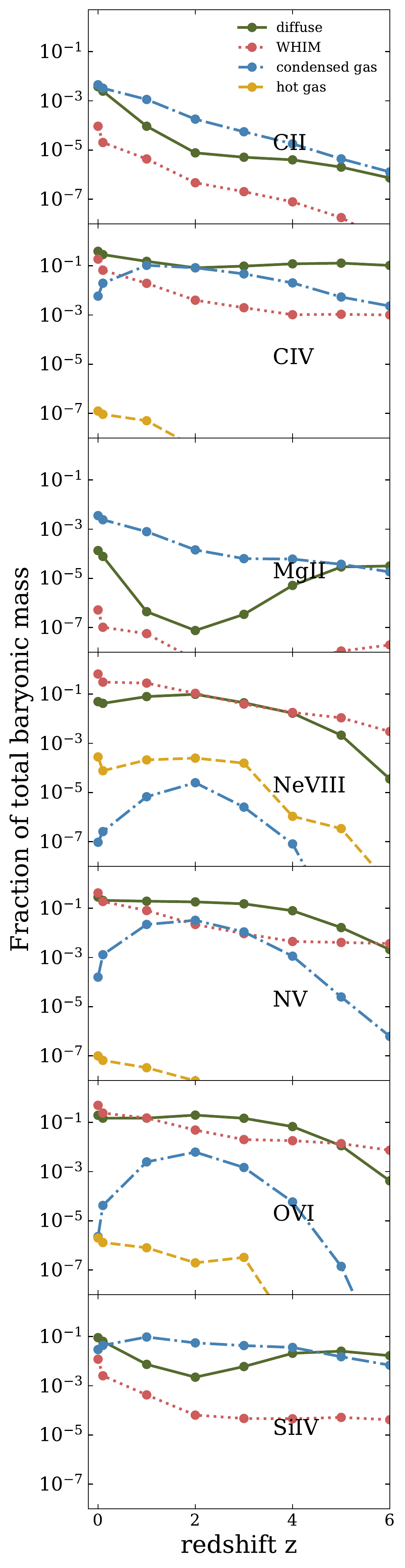}
 \end{subfigure}
 \begin{subfigure}{0.24\linewidth}
 \caption{\centering Haloes}
 \includegraphics[width=\textwidth]{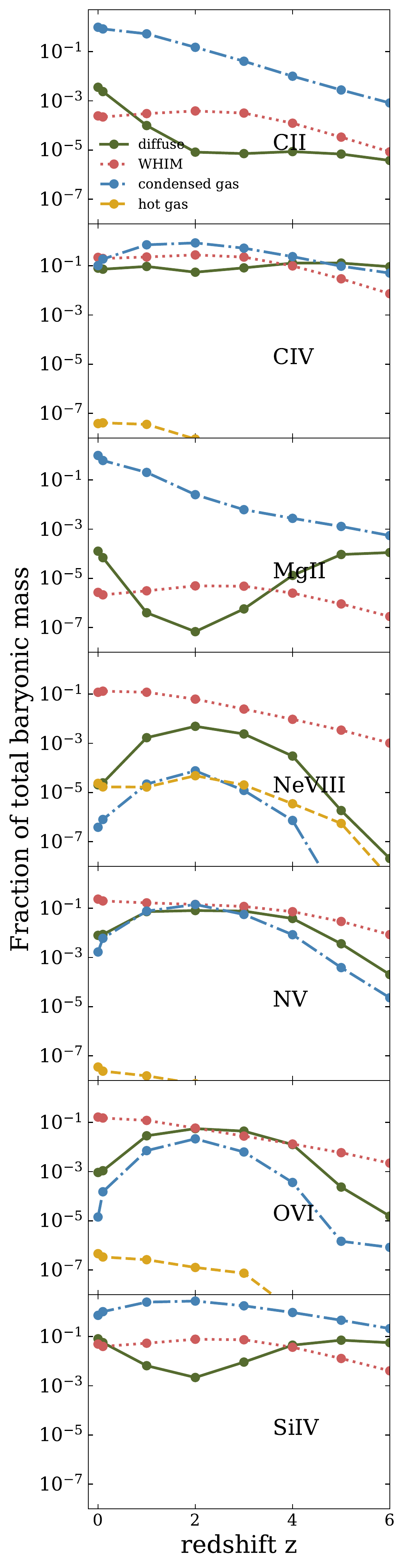}
 \end{subfigure}
\caption{Cosmic evolution of the mass fraction for the ionized metals normalized by the mass of each ion at $z=0$ (f$_{\rm ion}$ = M$_{\rm ionized\,metal,z}$/M$_{\rm ionized\,metal,z=0}$) split by gas phase in diffuse (green), WHIM (red), condensed gas (blue) and hot gas (yellow). We also divide the sample according the cosmic-web definition. From left to right: all the sample (voids+filaments+haloes), voids, filaments, and haloes.
Faint lines are used for condensed gas in voids to indicate that the fractions are computed with less than $10^4$ gas cells (see Section~\ref{sec:ions_phases} for further details). The results correspond to the TNG100 box.}
\label{fig:gasPhase_evol_ions}
\end{figure*}
%%%%%%%%%%%%%%%%%%%%%%%%%%%
%%%%%%%%%%%%%%%%%%%%%%%%%%%

We proceed now to investigate the distribution of ionized metals in the different gas phases. Figure~\ref{fig:T_rho_ions} shows the mass fraction of gas of the total baryonic mass in the density--temperature diagram for TNG100 at $z=0$, 0.1, 1, and 2. Vertical and horizontal lines demarcate the different gas-phase regions adopted.
We start by analysing the evolution of the diffuse phase, typically associated with the intergalactic low-dense and low-temperature gas. Our results indicate that this gas phase comprises 75.3 percent of the gas at $z=2$, a fraction that decreases significantly, to 37.3 per cent, at $z=0$. In addition, the comparison of the different panels shows that a large amount of diffuse cold gas increases its temperature with time, moving to the WHIM and hot phases. This is a consequence of the different feedback mechanisms and galactic outflows heating and ionizing the gas that reaches the ICM and IGM. 

Both the WHIM and hot-gas phases display an increase in the mass fraction with time. The WHIM phase makes up 17.8 per cent of the total gas at $z=2$, reaching 46.6 per cent at $z=0$. On the other hand, only 0.2 per cent of the gas is located in the hot phase at $z=2$, while this fraction raises to 6.8 per cent at $z=0$. The mass fractions for the different gas phases from $z=6$ to $z=0$ are listed in Table~\ref{table:mass_fraction}. We note that a comparison with previous works is difficult since the definitions of gas phases vary in the literature - the definition chosen for this work follows the one adopted in \citet{Haider2016}. Regardless of the definitions employed, our results are qualitatively in good agreement with previous findings  \citep[e.g.,][]{Rahmati2013,Oppenheimer2016}.

\begin{table*}
\caption{Mass fraction of baryons in the different gas phases at $z=0, 0.1, 1, 2, 3, 4, 5,$ and 6 for TNG100.}
\label{table:mass_fraction}
\centering
\begin{tabular}{lcccccccc}
\hline
            &  $z=0$ & $z=0.1$  & $z=1$ & $z=2$ & $z=3$ & $z=4$ & $z=5$ & $z=6$ \\
\hline
Condensed & 0.041 &  0.042  &  0.051 & 0.047 & 0.036 & 0.025 & 0.013 & 0.007 \\
Diffuse   & 0.373 & 0.387   &  0.564 & 0.753 & 0.863 & 0.923 & 0.961 & 0.977 \\
WHIM      & 0.466 & 0.463   &  0.335 & 0.178 & 0.091 & 0.047 & 0.024 & 0.014 \\
Hot       & 0.068 & 0.058   & 0.015  & 0.002 & 0.001 & 3.45$\times$10$^{-5}$ & 3.12$\times$10$^{-6}$ & 4.08$\times$10$^{-8}$\\
\hline
\end{tabular}
\end{table*}

The contours in Figure~\ref{fig:T_rho_ions} represent the regions enclosing 80 per cent of the total mass for each ionized metal. Particularly interesting in the context of the missing baryons problem is the evolution of Ne\,\textsc{viii}, O\,\textsc{vi}, and N\,\textsc{v}, which remain excellent tracers of the WHIM and diffuse gas throughout the redshift range considered. A fraction of these ions residing in diffuse gas at high redshift is in fact progressively pushed towards the WHIM gas phase by shock-heating and feedback processes that tend to increase gas temperature (this is particularly evident for Ne\,\textsc{viii}).
Also, a significant fraction of C\,\textsc{iv} is located in the diffuse and WHIM gas phases. As expected, Si\,\textsc{iv}, Mg\,\textsc{ii}, and C\,\textsc{ii} are, on the other hand, good tracers of the condensed gas up to high redshift.

Finally, Figure~\ref{fig:gasPhase_evol_ions} provides a complete summary of the main results presented in this paper regarding the cosmic evolution of ionized metals. The figure displays the mass fraction  normalized by the total mass of each ion at $z=0$ (f$_{\rm ion}$, defined as in Figure~\ref{fig:ions_VoidsFilamentsHaloes}) in condensed, diffuse, WHIM, and hot gas as a function of redshift and cosmic 
web environment. In the remainder of this section, we will describe the main features found.

For condensed gas, our results show that the abundance of C\,\textsc{iv},  Ne\,\textsc{viii}, N\,\textsc{v},  O\,\textsc{vi}, and  Si\,\textsc{iv}, typically increases from early on, peaking at $z\sim2$. The trends found for these ionized metals resemble the evolution of the cosmic star formation rate, and it can be explained as a direct consequence of the chemical enrichment from different feedback mechanisms together with gas cooling and metal recycling in dense regions \citep[][]{Pillepich2018}.  
On the other hand, C\,\textsc{ii} and Mg\,\textsc{ii} continue growing in mass up to $z=0$ (see Figure~\ref{fig:gasPhase_evol_ions}--a).

When the cosmic web environments are taken into account, we find that the mass fractions at $z=0$ in the condensed gas phase in haloes are 0.99 for C\,\textsc{ii} and Mg\,\textsc{ii}, while a value of 0.75 is measured for Si\,\textsc{iv} (Figure~\ref{fig:gasPhase_evol_ions}--d). Our results indicate that these metals are mostly in high-density and low-temperature regions. The mass fractions for the remaining ions in the condensed gas phase are distributed in haloes and filaments. 
As an example, C\,\textsc{iv} in condensed gas phase has a mass fraction of $\sim 0.1$ in haloes and $\sim0.005$ in filaments at $z=0$.

The connection between haloes and the condensed phase is easily understood, since both criteria map the densest and lowest-temperature regions of the Universe. These findings are in line with previous works. As an example, Mg\,\textsc{ii} has shown to be abundant in galactic haloes of luminous and massive galaxies, thus tracing the gas in transition between the IGM and the ISM. These are condensed and low-temperature gas regions, suggesting that  Mg\,\textsc{ii} is indeed a good tracer of star formation, galactic outflows, and accreted gas \citep[see e.g.,][]{Steidel2002,Zibetti2007,Weiner2009,Rubin2010,Martin2012,Rubin2012,Dutta2020,Anand2021}. It is also noteworthy that results obtained for the condensed phase should be treated with caution since numerical codes do not resolve properly the high-density low-temperature gas, due to a transition in sub-grid scaling for cooling and mass flow \citep[see][]{Chen2017,Vogelsberger2014}. Finally, we note that all the ionized metals in condensed gas in voids are modelled with less than $10^{4}$ cells over the entire TNG100 volume, resulting in a stochastic behavior for the trends. To indicate this caveat, we use blue faint lines for condensed gas in Figure~\ref{fig:gasPhase_evol_ions}--b.

A significant amount of C\,\textsc{iv}, N\,\textsc{v}, O\,\textsc{vi}, and Si\,\textsc{iv} is measured in a diffuse gas phase (f$_{\rm ion}$=0.48, 0.32, 0.23, 0.17 at $z=0$, respectively).
In particular, the mass fractions of N\,\textsc{v} and  O\,\textsc{vi} increase as redshift decreases.
Conversely, Mg\,\textsc{ii} and Si\,\textsc{iv} present a significant drop in their diffuse-gas mass fractions around $z\sim2$, followed by a subsequent increase (a factor of $\sim 10^{2}$ of difference for the mass fractions between these redshifts).
This might indicate that a fraction of the latter in diffuse gas is accreted to high-density regions, shifting to the cold phase at $z\sim2$ and progressively moving to low-dense regions due to stellar winds after the star formation peak. C\,\textsc{ii} shows a nearly constant mass fraction of the order of $\sim 10^{-5}$ for diffuse gas between $z=6$ and $z=2$, increasing towards low redshifts and reaching a fraction of $\sim 0.007$ at $z=0$ (Figure~\ref{fig:gasPhase_evol_ions}-a).

As for the WHIM phase, the mass fractions of C\,\textsc{iv}, Ne\,\textsc{viii}, N\,\textsc{v}, and O\,\textsc{vi} at $z=0$ are 0.41, 0.91, 0.68, and 0.77, respectively (Figure~\ref{fig:gasPhase_evol_ions}-a). If we also consider the cosmic web environment, we find that for C\,\textsc{iv} the mass fraction at $z=0$ is $\sim 0.22$ in haloes and $\sim0.19$ in filaments.
Most of Ne\,\textsc{viii} in WHIM phase is in filaments (f$_{\rm ion}=0.66$ at $z=0$), while the fraction in haloes is $\sim 0.12$  and in voids $\sim 0.13$. Similar mass fractions are also found for  N\,\textsc{v}, O\,\textsc{vi}. 
%O: 0.1 in voids, 0.16 haloes, 0.77
%N: 0.42 filaments, 0.23 haloes, 0.02 voids
Again, our results confirm that these ions are unique tracers of warm and low-density gas, where most of the missing baryons at low redshift are expected to be stored. On the other hand, the mass fractions of C\,\textsc{ii}, Mg\,\textsc{ii}, and Si\,\textsc{iv} in the WHIM phase are lower than those of the other ions considered. These latter ions present a subtle increase with redshift with a maximum fraction at around $z\sim2-3$ and a turn-down from $z\sim 2$ to $z=0$.

Finally, none of the ionized metals under analysis keep a large mass reservoir in the hot phase. Namely,  Ne\,\textsc{viii} has a mass fraction of $4\times10^{-4}$ at $z\sim 0$, while for 
O\,\textsc{vi} the measured fraction is $\sim 2.4\times10^{-6}$. The mass fractions are lower for C\,\textsc{iv} and N\,\textsc{v}, and no contribution is found for C\,\textsc{ii}, Mg\,\textsc{ii}, and Si\,\textsc{iv}. Overall, the mass budget in the hot gas phase increases towards low redshift.  

To summarize, we have shown that each ionized metal displays complex features, localizing its mass budget in different proportions across different gas phases and cosmic web environments. These findings are in agreement with previously reported numerical and observational measurements. Of particular importance in the context of the missing baryon problem at low redshift are our results for O\,\textsc{vi}, N\,\textsc{v}, and Ne\,\textsc{viii} as tracers of the WHIM and diffuse gas in filaments \citep[see e.g.,][]{Cen2006,Werner2008,Tepper-Garcia2011,Rahmati2013,Werk2014,Nelson2018_Oxigen,Roca-Fabrega2019}. In this context, \citet{Tepper-Garcia2011} investigated the physical properties of O\,\textsc{vi} absorbers at low redshift in the OWLS simulation. Their results suggest that the bulk of O\,\textsc{vi} mass is in gas with overdensities of $\rho/\langle \rho \rangle = 10^{0.5} - 10^{2.5}$, and temperatures in the range ${\rm T}=10^{4}-10^{6}{\rm K}$. Hence, detectable O\,\textsc{vi} absorbers are good tracers of shock-heated over-enriched gas. Also \citet{Marra2020} investigated the enriched gas in the CGM around galaxies from the EAGLE simulation, finding that ions such as O\,\textsc{vi} and Ne\,\textsc{viii} constitute a large fraction of gas with ${\rm T}\sim 10^{5.5}-10^{7}$~K in massive systems. \citet{Burchett2019} studied the CGM of 29 galaxies with masses M$_{*} \sim 10^{9}-10^{11} {\rm M}_{\odot}$ at $0.49<z<1.44$. Their results suggest that Ne\,\textsc{viii} in CGM gas encompasses $\sim 6-20$\% of the total baryonic mass. Also \citet{Tejos2016} studied the gas in filaments connecting massive galaxy cluster pairs, showing that warm-hot gas ($10^5-10^6$~K) is enhanced within filaments.

In conclusion, the fact that O\,\textsc{vi}, N\,\textsc{v}, and Ne\,\textsc{viii} are located in filaments in the WHIM and diffuse gas phases, together with the fact that a significant baryonic mass budget is in these phases at $z=0$  ($>70\%$), adds up to the growing evidence that these ionized metals are good tracers of the missing baryons.

\section{Summary and conclusions}\label{sec:conclusions}

Understanding the evolution and distribution of metals on large scales provides information about the processes that regulate star formation and gas dynamics within and around galaxies. Moreover, studying these distributions allow us to map the baryons on large scales and thus shed light on the so-called missing baryon problem discussed at low redshift.
Current state-of-the-art hydrodynamical cosmological simulations have improved the way they model and reproduce different observational tracers, providing an exceptional tool to track the evolution of baryons across cosmic time \citep[see e.g.,][]{Cen1999,Cen2006,Cen2011,Tepper-Garcia2011,Oppenheimer2012,Oppenheimer2016,Nelson2018_Oxigen,Martizzi2018,Roca-Fabrega2019,Marra2020,Martizzi2020,Strawn2020}. Despite this success, even the most accurate simulations producing self-consistent outflows also need calibration of their sub-grid models. This, in turn, impacts the hydrodynamics of the gas and distribution of metals. Different works have indeed confirmed the impact that sub-grid models have on the temperature distribution of the gas and its ionization state \citep[see e.g.,][]{Cen2011,Roncarelli2012,Oppenheimer2016}. Hence, caution must be exercised when comparing observational results with those obtained from hydrodynamical simulations.

    In this work, we investigate the large-scale distribution of ionized metals in the IllustrisTNG simulation across cosmic time, from $z=6$ to $z=0$. To this end, we use two different simulation boxes with the highest resolution available for each size in the IllustrisTNG suite: TNG100-1, with side length 75~$h^{-1}$~Mpc and  TNG300-1, spanning 205~$h^{-1}$~Mpc. The IllustrisTNG suite represents a revised version of the Illustris simulation, which among other features, reproduces the cosmic star formation rate, the galaxy stellar mass function, and includes updated prescriptions for the chemical yields -- all fundamental elements for this work.

With the aim of providing a revised vision of the missing baryon problem, we map the redshift evolution of seven ionized metals, C\,\textsc{ii}, C\,\textsc{iv}, N\,\textsc{v}, Ne\,\textsc{viii}, Mg\,\textsc{ii}, O\,\textsc{vi}, and Si\,\textsc{iv}, in the cosmic web (haloes, filaments, and voids), and in the different gas phases (WHIM, hot, diffuse, and condensed gas). The selected set of ionized metals is chosen because they are commonly detected in observations. Moreover, they are excellent tracers of the baryonic content on large scales, mapping the cosmic web and the different gas phases. 
Our main results can be summarized as follows:
\begin{itemize}
    \item We find that the mass fraction of baryons in WHIM and hot-gas phases (with ${\rm T}>10^5$K) increases significantly as the redshift decreases. Namely, the WHIM phase represents 17.8 per cent of the gas at $z=2$, reaching 46.6  per cent at $z=0$. These trends are a consequence of galactic outflows heating and shifting the gas from high-density to low-density regions in the ICM and IGM (Figure~\ref{fig:T_rho_ions}).
    \item We show that each ionized metal maps different parts of the density-temperature diagram and cosmic-web environments. Furthermore, their total masses and distributions evolve with redshift as a consequence of chemical enrichment due to star formation, and AGN and stellar feedback processes together with strong kinetic outflows expelling enriched gas to the outer parts of the galaxy in the IGM.  
    \item We find that most of the mass budget for C\,\textsc{ii} and Mg\,\textsc{ii} is in haloes in the condensed gas phase (at $z=0$ it represents 99 per cent of the mass fraction of each of these ions). Our findings suggest that C\,\textsc{ii} and Mg\,\textsc{ii} are excellent tracers of star-forming regions and the gas in transition between the IGM and the ISM (Figure~\ref{fig:ions_VoidsFilamentsHaloes}).
  \item 
 Our findings show that C\,\textsc{iv} and Si\,\textsc{iv} present similar evolution in their mass fractions.
 For instance, the mass budget in condensed gas in haloes resembles the cosmic star formation rate evolution, with a maximum around $z\sim2$ (Figure~\ref{fig:gasPhase_evol_ions}-d). This is explained by chemical enrichment from stellar feedback combined with gas cooling in dense star-forming regions. 
 At $z=0$, the mass fraction for C\,\textsc{iv} and Si\,\textsc{iv} in condensed gas inside haloes is $\sim 0.10$ and $\sim0.73$, respectively. The diffuse and WHIM-phase components of these ions show similar trends.
 \item For O\,\textsc{vi}, N\,\textsc{v}, and Ne\,\textsc{viii} most of the mass budget is in the WHIM phase in filaments. The mass fractions in these regions increase as redshift decrease. In particular, for Ne\,\textsc{viii}, we find a mass fraction of $\sim 0.91$ in the WHIM at $z=0$. When analysing different cosmic web environments separated, the mass fraction measured in filaments is $\sim0.66$, while values of $\sim0.12$ and 
$\sim 0.13$ are obtained in haloes and voids, respectively.
Similar results are found for N\,\textsc{v} and O\,\textsc{vi} (Figure~\ref{fig:gasPhase_evol_ions}). These findings suggest that Ne\,\textsc{viii}, N\,\textsc{v}, and O\,\textsc{vi} are the best tracers of the low-density and warm/hot gas.
\end{itemize}

Our results confirm that most baryons at low redshift are in low-density regions and in the form of warm/hot gas. Among the different ionized metals considered, Ne\,\textsc{viii},  N\,\textsc{v}, and O\,\textsc{vi} in filamentary structure emerge as the best tracers of this baryonic component. We expect that the predictions reported in this work can contribute to the quest of unveiling the long-standing mystery of the missing baryons in the local Universe. In this context, our results will hopefully serve as an additional guide for upcoming surveys aimed at scanning the cosmic web through X-ray absorption spectroscopy with instruments such as the Athena Observatory.

\section*{Acknowledgements}
We thank the anonymous referee for her/his careful reading and constructive comments which helped improve the paper. MCA acknowledges Dylan Nelson, and Annalisa Pillepich for useful comments. MCA acknowledges financial support from the Austrian National Science Foundation through FWF stand-alone grant P31154-N27. ADMD thanks FAPESP for financial support. AMD also thanks Fondecyt for financial support through the Fondecyt Regular 2021 grant 1210612.
SB was supported by NSF grant AST-1817256. FM acknowledges support through the Program "Rita Levi Montalcini" of the Italian MUR. The computations in this paper were run on the FASRC Odyssey cluster supported by the FAS Division of Science Research Computing Group at Harvard University.

\section*{Data availability}
The data underlying this article will be shared on reasonable request to the corresponding authors.

\bibliographystyle{mnras}
\bibliography{draft}

\label{lastpage}

\appendix

\section{Cosmic evolution of baryons in the cosmic web}

In this appendix we present the evolution of the mass fraction of baryons according to the cosmic web definition adopted in this work. For a further discussion we refer to \citet{Martizzi2018}, while the results obtained with a similar definition using Illustris simulations can be found in \citet{Haider2016}.

\begin{figure}
\centering
\includegraphics[width=0.5\textwidth]{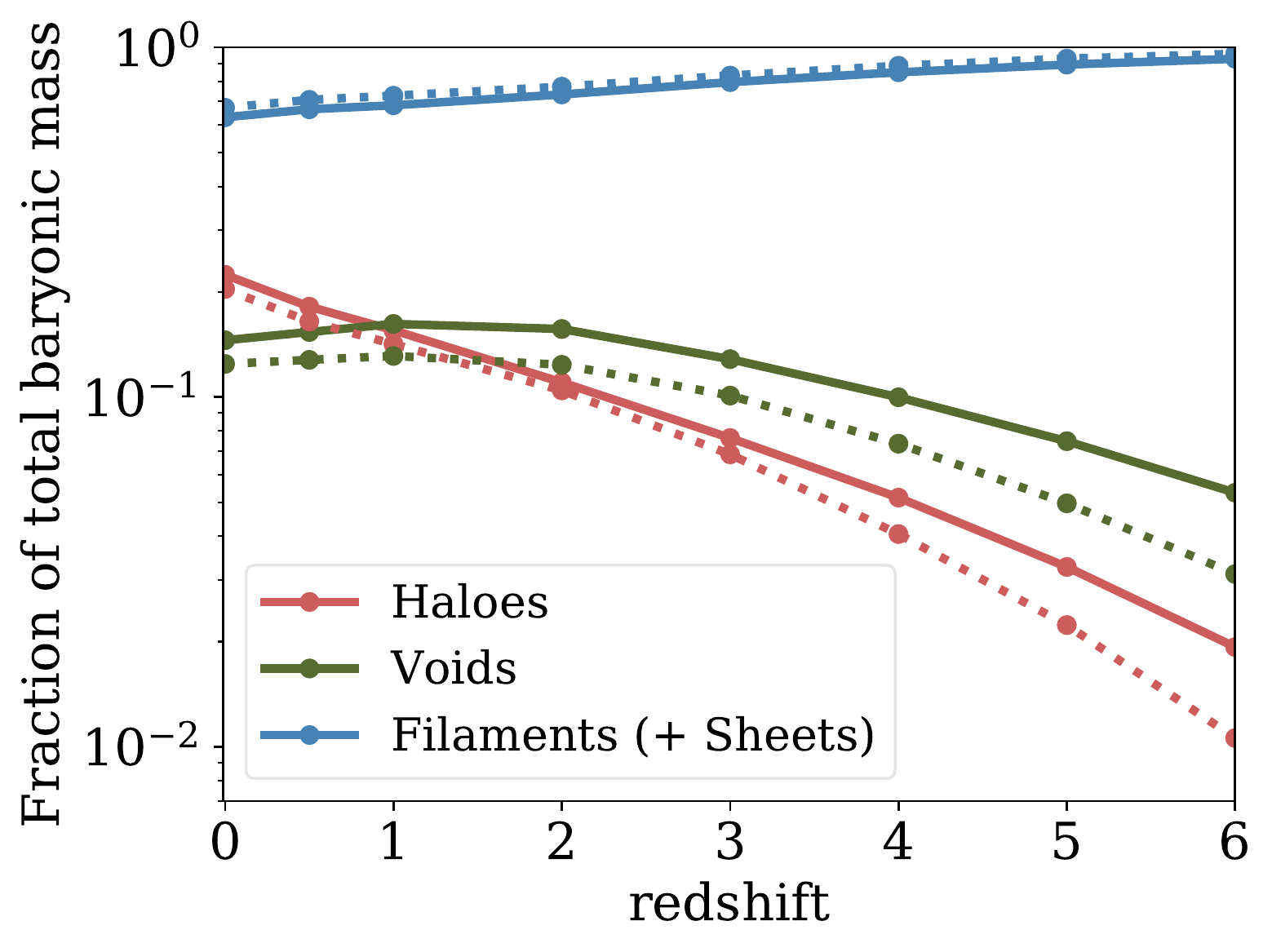}
\caption{Evolution of the total baryonic mass fraction of the cosmic structures defined in this work from redshift $z=6$ to $z=0$ for TNG100 (solid lines) and TNG300 (dotted lines). As stated in \citet{Martizzi2018}, we note that the way we define the filaments corresponds to the definition of filaments and sheets when computing the cosmic structures with the tidal tensor.}
\label{fig:Appendix_CosmicStructures}
\end{figure}

Figure~\ref{fig:Appendix_CosmicStructures} shows the baryonic mass fraction in voids, filaments, and haloes as a function of redshift for TNG100, and TNG300. 
Our results show that the mass fraction in filaments decreases as the redshift decreases. 
This is explained since our method, based on the local dark matter density, encloses filaments and sheets in a single definition.
Hence, as discussed in \citet{Martizzi2018}, while for other approaches based on tidal tensor the mass fraction in filaments increases with redshift, our results reflect the contribution of filaments and sheets together.  

We also find that the mass fraction in voids increases from $z=6$ to $z=2$. As stated in \citet{Haider2016}, at high redshift the dark matter is distributed homogeneously, with a density $\sim \Omega_{dm}$\rhocrit. Hence, most of the mass falls in the filament category, and the under-dense regions are only created when the matter from less dense regions is pulled into denser regions as the redshift decreases.

%%%%%%%%%%%%%%%%%%%%%%%%%%%%%
%%%%%%%%%  FIG A1  %%%%%%%%%%
%%%%%%%%%%%%%%%%%%%%%%%%%%%
%%%%%%%%%%%%%%%%%%%%%%%%%%%

\end{document}